\documentclass[preprint,12pt]{elsarticle}

\usepackage{amssymb}
\usepackage{amsmath}

\usepackage{graphicx} 
\usepackage{caption} 
\usepackage{subcaption} 
\usepackage{float} 
\usepackage{epstopdf} 
\DeclareGraphicsExtensions{.pdf,.png,.jpg,.eps} 
\usepackage{hyphenat} 
\usepackage{soul} 
\usepackage{amsmath} 
\usepackage{amssymb} 
\usepackage{cancel} 
\usepackage{multirow} 
\usepackage[utf8]{inputenc} 
\usepackage[T1]{fontenc} 
\usepackage[margin=1in]{geometry} 
\usepackage{pdflscape} 
\usepackage{titlesec} 
\usepackage{tikz} 
\usepackage{arydshln} 
\usepackage{placeins} 
\usepackage{xcolor} 
\usepackage[acronym,shortcuts]{glossaries} 
\makeglossaries 
\newacronym{GFVSC}{\textcolor{black}{GFM-VSC}}{Grid-forming voltage source converter} 

\journal{International Journal of Electrical Power \& Energy Systems}

\begin{document}

\begin{frontmatter}



\title{Impact on transient stability of self-synchronisation control strategies in grid-forming VSC-based generators\color{black}}

\author[label1,label2]{R\'egulo E. \'Avila-Mart\'inez}
\author[label3]{Xavier Guillaud}
\author[label2]{Javier Renedo}
\author[label1,label2]{Luis Rouco}
\author[label1,label2]{Aurelio Garcia-Cerrada}
\author[label1,label2]{Lukas Sigrist}
\affiliation[label1]{organization={Instituto de Investigación Tecnológica (IIT)},
            addressline={C. del Rey Francisco, 4},
            postcode={28008},
            state={Madrid},
            country={Spain}}

\affiliation[label2]{organization={ETSI ICAI, Universidad Pontificia Comillas},
            addressline={Alberto Aguilera, 23},
            postcode={28015},
            state={Madrid},
            country={Spain}}

\affiliation[label3]{organization={Laboratoire d'Electrotechnique et d'Electronique de Puissance (L2EP), Ecole Centrale de Lille},
            addressline={Cité Scientifique, CS 20048},
            city={Villeneuve d'Ascq},
            postcode={59651},
            state={Hauts-de-France},
            country={France}}


\begin{abstract}
Voltage source converters with grid-forming control (GFM-VSC) are emerging as a solution for integrating renewable energy resources (RERs) into power systems. GFM-VSCs need a self-synchronisation strategy to ensure that all converters and generators in the power system are in synchronism and they reach \color{black} the same frequency in steady state. The self-synchronisation strategy in GFM-VSCs that has received most attention in previous research is virtual synchronous machine (VSM) control scheme. However, no systematic study of the effects on transient stability of different variants of GFM-VSC self-synchronisation strategies has been carried out in previous work. \color{black}This paper analyses and compares transient stability of four \color{black} self-synchronisation strategies for \acp{GFVSC}: VSM without phase-locked loop (PLL), VSM with PLL, VSM without PLL using wash-out filter and integral-proportional (IP) controller. The paper also analyses two \color{black} different methods that can \color{black} be applied to \ac{GFVSC} self-synchronisation strategies to improve transient stability: the concept of virtual unsaturated active-power controller (VAPC), proposed in previous work, and an algorithm for frequency limitation in the \acp{GFVSC} (FLC), which is proposed in this paper.
\end{abstract}




\end{frontmatter}




This is an unabridged draft of the following paper submitted to the International Journal of Electrical Power \& Energy Systems on September 1st 2025:
\begin{itemize}
\item R. E. Ávila-Martínez, X. Guillaud, J. Renedo, L. Rouco, A. Garcia-Cerrada, L. Sigrist, "Impact on transient stability of self-synchronisation control strategies in grid-forming VSC-based generators", submitted to the International Journal of Electrical Power \& Energy Systems, pp. 1-36, September 2025.
\item Internal reference: IIT-25-279WP.
\end{itemize}

\newpage

\section{Introduction}\label{sec:sscs_intro}
\noindent Voltage source converters with grid-forming control (\acp{GFVSC}, for short) are emerging as the preferred \color{black} solution for integrating renewable energy resources (RERs) into power systems~\cite{Paolone2020,Strunz2023,Carmen_Cardozo2024}. \acp{GFVSC} require self\hyp{}synchronisation control strategies, whose main philosophy is to mimic \color{black} the behaviour of conventional synchronous machines. Self-synchronisation strategies are necessary to ensure that all \ac{GFVSC} converters and synchronous generators in the system reach the same steady-state frequency \cite{Barker2021}\color{black}. Unlike for grid\hyp{}following VSCs (GFL-VSC), a phase\hyp{}locked loop (PLL) is not required for synchronisation in \acp{GFVSC}, and self-synchronisation strategies can be used instead.

Previous studies have proposed several variants of self-synchronisation strategies in \acp{GFVSC}, such as power synchronisation control~\cite{Zhang2010}, synchronverter control~\cite{Zhong2011}, virtual synchronous machine (VSM) control~\cite{DArco2014,DArco2015,jroldan2019}, active\hyp{}power/frequency (P-f) droop control \cite{Rocabert2012}, integral-proportional (IP) controller \cite{Qoria2020}, virtual oscillator \cite{jouini2018, GroßD2019} and reactive\hyp{}power\hyp{}based synchronisation\color{black}~\cite{Amenedo2021}, among others. Authors in~\cite{Mandrile2023} analysed some of those different self-synchronisation strategies and proposed a variant to improve the dynamic response of the \ac{GFVSC} by using a lead/lag filter\color{black}. Reference~\cite{JaumeGB2024} analysed the performance of different \ac{GFVSC} self-synchronisation algorithms taking into account the primary energy source of the \ac{GFVSC} power converter, and the work proposes a novel control scheme taking this aspect into account (so-called resource-aware grid-forming (RA-GFM)), improving the performance of the system.  A summary of \ac{GFVSC} strategies (in this case applied to high voltage direct current systems based on voltage source converters (VSC-HVDC)) can be found in~\cite{FG_Puricelli_phd2025} and \ac{GFVSC} control applied to VSC-HVDC systems has been analysed in~\cite{FG_Puricelli2025}.   \color{black}

VSM self-sycnhronisation control has attracted considerable attention in power systems research and several variants have been described, including\color{black}: VSM without phase-locked loop (PLL)~\cite{DArco2014}, VSM with PLL~\cite{DArco2015,jroldan2019} and VSM without PLL using a wash-out filter~\cite{MoO_DArco2017}. The work in \cite{DArco2014} 
proved analytically that the known P-f-droop and VSM-without-PLL self-synchronisation strategies are equivalent. 

It is essential to fully understand and ensure power system stability of power systems with large amounts of power converters, including \acp{GFVSC}. \color{black} Power system transient stability is defined as large-disturbance angle stability and it is the ability of the generators of the system to remain in synchronism and reach a new stable steady-state equilibrium point when subject to large disturbances \cite{Nikos2021}. Although this phenomenon was originally related to conventional synchronous machines, research publications during the past years have shown that loss-of-synchronism phenomenon is also present in \acp{GFVSC}~\cite{Andrade2011,Shuai2018,Cheng2020,Pan2020,Qoria_VSC_CCT2020}. Furthermore, the work in \cite{Ledesma2025} analysed the impact of \ac{GFVSC} and GFL-VSC power converters on angle stability under lager disturbances in large-scale power systems, concluding that the former present better behaviour in terms of transient stability. \color{black} 

Different publications have analysed the transient stability of power systems with \acp{GFVSC} using different self-synchronisation strategies. Several studies~\cite{Andrade2011,Shuai2018,Cheng2020,Pan2020,Qoria_VSC_current_limit2020,Choopani2020,BlaabjergFTSAngle2022,XiongX_2021a,RAvilaM2022,SiW2023} used VSM without PLL control (or P-f droop control) while another~\cite{Qoria2020} proposed the IP controller. The latter study 
also proved the equivalence between the IP controller and VSM-with-PLL scheme under small disturbances; however, the performance of the latter under large disturbances was not analysed. Reference \cite{Luo2025} compared the performance of VSM-without-PLL and VSM-with-PLL schemes in terms of small-signal stability, but transient-stability performance was not analysed. \color{black} Furthermore, researchers in \cite{BlaabjergFTSAngle2022} proposed the 
Transient Damping Method (TDM), 
a supplementary active-power set point (containing a proportional gain and a wash-out filter) to improve transient stability. 
This study was performed in VSM without PLL control. 
Although the configuration when using the TDM is similar to self-synchronisation strategy VSM without PLL control + wash-out, it is not exactly the same. To the best of the authors' knowledge, previous publications have not analysed the transient stability phenomenon in power systems with \acp{GFVSC} using VSM without PLL and with washout as self-synchronisation strategy. 

Although the transient-stability performance of several \ac{GFVSC} self-synchronisation strategies has been studied separately, a comparison of each of them has not been carried out in previous work. Hence, a comprehensive understanding of the transient-stability behaviour and the advantages and disadvantages of each of these strategies has not yet been reached. 

This paper addresses this gap by providing a comprehensive analysis and comparison of the transient-stability performance of different \ac{GFVSC} self-synchronisation control strategies, including VSM control (without and with PLL), VSM control without PLL and with wash-out and the IP controller technique. 
It analyses the advantages, limitations and performance characteristics of each self-synchronisation control strategy under different operating conditions in terms of transient stability.

A recent work~\cite{VattaKkuniK2024} proposed the concept of virtual active-power control (VAPC) to improve the transient-stability performance of \acp{GFVSC} using a variant self-synchronisation \color{black} strategy similar to VSM without PLL control, but with an additional degree of freedom. 
Later studies~\cite{LabaY2023,Yorgo2025} applied the concept of VAPC to \acp{GFVSC} using IP-control. Furthermore, reference~\cite{Camboni2025} analysed transient stability of \acp{GFVSC} when using VSM without PLL and VAPC, proving the effectiveness of VAPC concept when using different priorities of $d-q$-axis currents in the current saturation algorithm of the power converter.  However, the concept of VAPC has not been applied to \ac{GFVSC} strategies VSM - without PLL or VSM - with washout explicitly in previous publications. Hence, questions that arise are whether the concept of VAPC is effective when using different self-synchronisation strategies and the comparison of the effectiveness of VAPC concept when using different \ac{GFVSC} strategies. \color{black} This paper addresses this gap and analyses systematically the impact of VAPC to the transient-stability performance of the most extended \ac{GFVSC} self-synchronisation strategies. 

The loss-of-synchronism phenomenon in \acp{GFVSC} occurs due to significant variations of the angle of the modulated voltage of the \ac{GFVSC}, producing, therefore, significant variation on its frequency. Some methods applied to self-synchronisation strategies in \acp{GFVSC} seek to prevent loss of synchronism, by limiting directly or indirectly the output frequency of the \ac{GFVSC}. The work in~\cite{Qoria2020} proposed increasing the virtual inertia of the converter during the fault to improve transient stability. The inertia increase is activated when the current injection of the \ac{GFVSC} reaches its limit.   Reference~\cite{Collados-Rodriguez_2023} proposed an adaptive P-f droop-based control method, which leads to an indirect limitation of the \ac{GFVSC} output frequency, preventing loss of synchronism. The additional adaptive P-f droop term is introduced when the frequency of the \ac{GFVSC} goes outside pre-defined limits. Research studies in~\cite{Du2019,Du2024, Du2025} proposed a scheme for active-power/frequency limitation in \acp{GFVSC}, which consisted in limiting the active-power injection of the converter by changing its output frequency and limiting it with a saturator, improving both transient stability and frequency stability. The active-power/frequency limitation process is triggered when the active-power injection of the \ac{GFVSC} goes outside its limits. Although main application of the frequency-limitation method proposed in~\cite{Du2019,Du2024} is frequency stability and limiting the \ac{GFVSC} active-power injection when subject to power imbalances, it also helps to improve transient stability. The works in~\cite{XZhao_2020,XZhao_2022} proposed a frequency freezing strategy in \acp{GFVSC} to improve transient stability. When a the current injection of the \ac{GFVSC} reaches its limit, the output frequency of the \ac{GFVSC} is freezed to the pre-fault frequency, preventing loss of synchronism.  Finally, the work in~\cite{Saman_2022} and~\cite{Li2024} proposed a frequency saturation method in \acp{GFVSC}: a saturator is added to the output frequency of the \ac{GFVSC}, which is activated if the frequency goes outside pre-defined limits. The work in~\cite{Li2024} includes the additional condition that the derivative of the frequency has the same sign than the frequency. The work in~\cite{Li2024} proposes the frequency saturator to improve transient stability explicitly, while the work in~\cite{Saman_2022} just focus the study in fault-ride-through (FRT) performance of the \ac{GFVSC}, but the paper does not analyse loss-of-synchronism phenomenon. Reference~\cite{Li2024} proposes to tune the frequency limits of each \ac{GFVSC} using using data of the whole power system (e.g., virtual inertia of all the \acp{GFVSC} of the system), to improve transient stability. 

While references~\cite{Qoria2020,Collados-Rodriguez_2023,Du2019,Du2024,XZhao_2020,XZhao_2022,Saman_2022,Li2024} clearly prove the potential of frequency-limitation methods in \acp{GFVSC} to improve transient stability, there is still room for further improvement. For example, references~\cite{Qoria2020} and~\cite{Collados-Rodriguez_2023} limit the output frequency of the \ac{GFVSC} indirectly; further improvements in terms of transient stability could be achieved if the output frequency of the \ac{GFVSC} were limited directly during the fault. The work in~\cite{XZhao_2020,XZhao_2022} limits the output frequency of the \ac{GFVSC} directly but it freezes the value to the pre-fault frequency. It would be desirable to maintain the \ac{GFVSC} self-synchronisation behaviour as much as possible during the fault. Hence, a frequency saturator (as in~\cite{Du2019,Du2024,Saman_2022,Li2024}) rather than a freezed frequency may be more practical. Regarding the activation of the frequency limiter, in references~\cite{Qoria2020,XZhao_2020,XZhao_2022} the frequency limitation process is activated when the current injection of the \ac{GFVSC} reaches its limit. Naturally, this occurs when a fault close to the converter occurs. However, the current limit could also be reached due to changes in the operating conditions or other disturbances, where the frequency limitation process may not be desirable or needed. In~\cite{Collados-Rodriguez_2023,Saman_2022,Li2024}, the frequency limitation process is activated when the output frequency of the \ac{GFVSC} reaches pre-defined limits, which may not distinguish if the event is related to a transient-stability or a frequency-stability issue. For example, if there is a power imbalance in the system the frequency of the centre of inertia of the system, the bus frequencies and the frequencies of the \acp{GFVSC} will change, but this does not necessarily mean that there is a transient stability issue: the frequencies of all \acp{GFVSC} of the system may be in synchronism during the transient. In fact, frequency excursions in a relatively small power system could be large due to power imbalances, but a single \ac{GFVSC} could loss synchronism caused by a nearby fault, and at this moment and the frequency increment of the \ac{GFVSC} could be much smaller than admissible frequency limits of the power system and the frequency of the \ac{GFVSC} will be different than the frequencies of other generators, during the transient. The same issue occurs with the proposal of~\cite{Du2019,Du2024}, which is activated when the active power injection of the \ac{GFVSC} goes outside its limits.  

Motivated by these publications (\cite{Qoria2020,Collados-Rodriguez_2023,Du2019,Du2024,XZhao_2020,XZhao_2022,Saman_2022,Li2024}), this paper proposes an algorithm for internal \color{black} frequency limitation in \ac{GFVSC} self-synchronisation strategies to improve transient stability (FLC, for short). In fact, some of the ideas of each strategy proposed in previous work \color{black} are combined in such way that the advantages some of the strategies are used together, while some new features are also proposed here. Details will be provided in Subsection~\ref{sec:sscs_freq_lim}. \color{black} The frequency limiter proposed in this work (FLC) is applied directly  and it is only activated during the fault, which is detected with an voltage-based hysteresis characteristic, ensuring an explicit limitation in the frequency imposed by the \ac{GFVSC}. In addition, the frequency limits are calculated around the pre-fault frequency, to avoid potential issues due to pre-fault steady-state frequency deviations. Hence, the proposed frequency limiter has advantages from a practical point of view. Finally, the frequency-limitation methods proposed in~\cite{Qoria2020,Collados-Rodriguez_2023,Du2019,Du2024,XZhao_2020,XZhao_2022,Saman_2022,Li2024} are analysed for particular \ac{GFVSC} self-synchronisation schemes, while this work analyses the proposed frequency limiter for different \ac{GFVSC} self-synchronisation strategies. 

Along the lines described above, the scientific contributions of this paper can be summarised as follows:
\begin{itemize}
    \item Comprehensive analysis and comparison of transient-stability performance of different \ac{GFVSC} self\hyp{}synchronisation control strategies: VSM control without and with PLL, IP controller and the VSM control without PLL using a wash-out filter.
    \item Analysis of the impact of the concept of VAPC to transient-stability performance of the the studied \ac{GFVSC} self\hyp{}synchronisation strategies.
    \item Proposal of a simple but effective algorithm for frequency limitation (FLC) \color{black} in self-synchronisation strategies in \acp{GFVSC} to improve transient stability. 
    \item Analysis the impact on transient stability of the wash-out-filter time constant in VSM- with-washout \ac{GFVSC} strategy. \color{black}
\end{itemize}
The analysis is carried out by ElectroMagnetic Transient (EMT) simulation in Matlab + Simulink + SimPowerSystems, using detailed average-type EMT models of power converters, including their controllers and operating limits.

The rest of the paper is organised as follows. Section~\ref{sec:VSC_V} describes modelling and control aspects of \acp{GFVSC}, following the approach used in this paper. Section~\ref{sec:sscs_VSM} presents the \ac{GFVSC} self-synchronisation strategies that will be analysed and compared in the paper.  Section~\ref{sec:sscs_VAPC_FL} describes two methods to improve transient stability of \acp{GFVSC}, the concept of VAPC and the frequency limiter (FLC) proposed in this paper. Section~\ref{sec:Theoretical_analysis} presents a theoretical analysis of transient-stability performance of the \ac{GFVSC} self-synchronisation strategies and the concepts of VAPC and FLC. The case study and simulation results are presented in Section~\ref{sec:sscs_Results_VSM_IP}, including time-domain simulation and calculation of critical clearing times (CCT). Section~\ref{sec:sscs_conclusion} presents the conclusions of the paper. Finally, data used in the paper can be found in the Appendix.    
\color{black}

\newpage

\section{Modelling and control of GFM-VSCs}\label{sec:VSC_V}
\begin{figure}[htbp]
\centering
\includegraphics[width=0.8\textwidth]{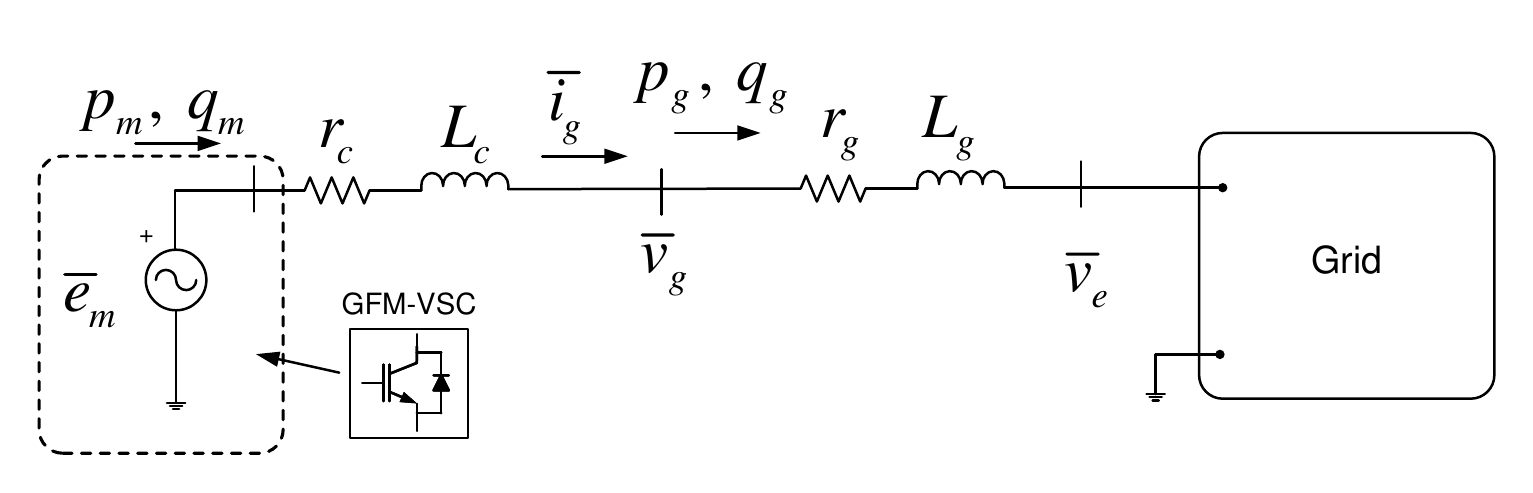}
\caption{Equivalent model of a Grid-Forming Voltage Source Converter (\ac{GFVSC}) connected to the grid.}
\label{fig:VSC_V_model_sscs}
\end{figure}

\noindent  Fig.~\ref{fig:VSC_V_model_sscs} shows the scheme of a \ac{GFVSC} connected to an infinite grid. $\bar {e}_{m}$ is the modulated voltage of the \ac{GFVSC}, and $\bar{z}_{c}=r_{c}+j x_{c}$ ($x_{c}=L_{c} \omega$) is its series connection impedance, which includes its equivalent series reactor and the impedance of the connection transformer. The infinite grid has a voltage $\bar {v}_{e,i}$ and an equivalent impedance $\bar{z}_{g}=r_{g}+j x_{g}$. The voltage at Point of Common Coupling (PCC) of the \ac{GFVSC} is $\bar{v}_g$ and the current and active- and reactive-power injections of the \ac{GFVSC} at the PCC are $\bar{i}_g$, $p_g$ and $q_g$, respectively.  

\begin{figure}[!htbp]
\begin{center}
\includegraphics[width=1\columnwidth]{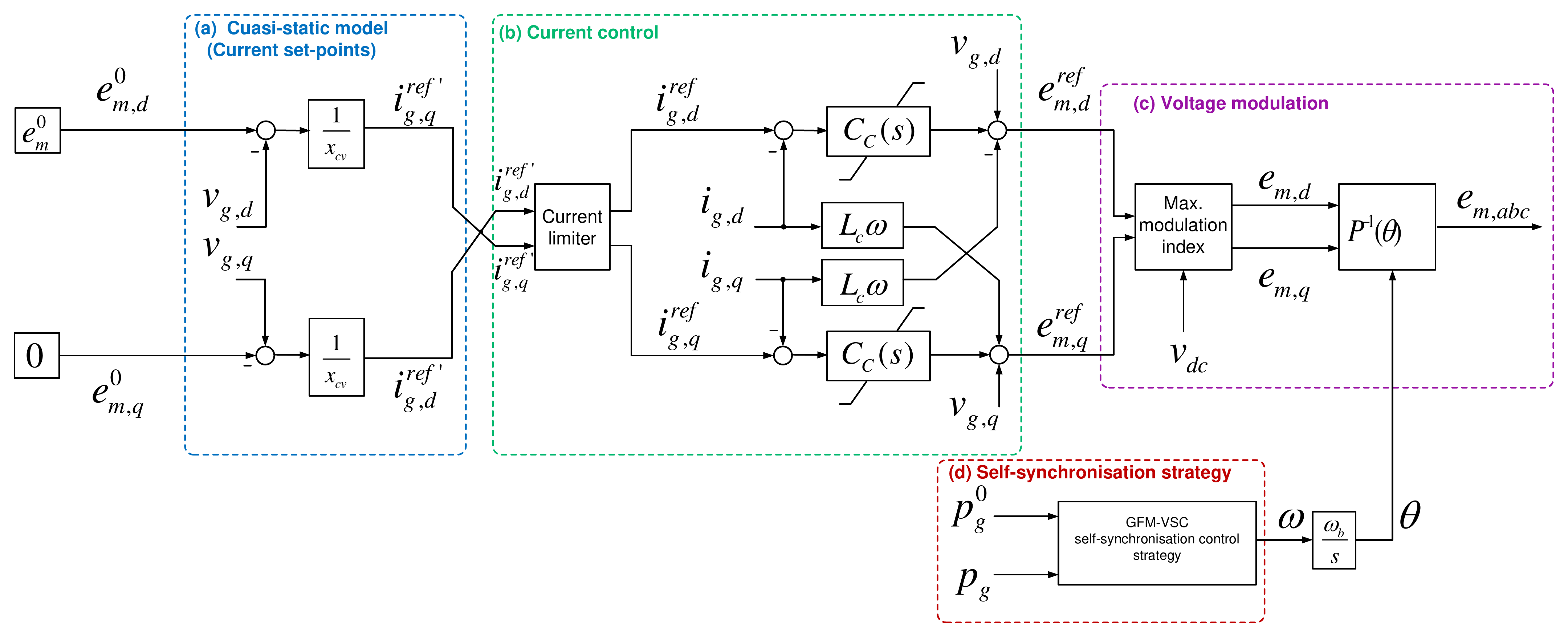}
\caption{General scheme of the control system of a GFM-VSC. }
\label{fig:VSC_L_control_loops_general}
\end{center}
\end{figure}

Fig.~\ref{fig:VSC_L_control_loops_general} shows a general scheme of the control system of a GFM-VSC (see \cite{VattaKkuniK2023,LabaY2023} for example). This work uses a current-controlled \ac{GFVSC} approach. \color{black} The \ac{GFVSC} controller is based on vector control using a mobile $d-q$ reference frame, where the modulated voltage $\bar{e}_m$ is aligned with mobile $d-q$ axes: $e_{m,d}=e_m$ and $e_{m,q}=0$. The \ac{GFVSC} has a constant set point for the magnitude of the modulated voltage ($e_{m}^0$), and it uses a quasi-static model to calculate the current set points (marked with (a) in Fig.~\ref{fig:VSC_L_control_loops_general}), following the approach of~\cite{MoO_DArco2017,VattaKkuniK2023,LabaY2023}: \color{black}
\begin{eqnarray}\label{current_setpoints_2}
	i_{g,q}^{ref'} = \frac{{v}_{g,d} - e_{m,d}^{0}}{x_{cv}}, \mbox{\space}
	i_{g,d}^{ref'} = \frac{e_{m,q}^{0} - {v}_{g,q}}{x_{cv}}. 
\end{eqnarray}
where $e_{m,d}^{0}=e_{m}^{0}$ and $e_{m,q}^{0}=0$ and $x_{cv}=x_{c}+x_{v}$ is an equivalent virtual reactance used for current-setpoint calculation. The equivalent virtual reactance ($x_{cv}$) contains the true (or estimated) series reactance ($x_{c}$) and an additional virtual reactance ($x_{v}$). Notice that $x_{v}=0$ pu could be used, or $x_{v}>0$ pu if required to limit active-power variation in case of a phase shift.\color{black}

A current controller is implemented ((b) in in Fig.~\ref{fig:VSC_L_control_loops_general}) with Proportional-Integral (PI) controllers and with a current limiter. Conventional Current Saturation Algorithm (CSA) is used for current limitation \cite{Qoria_VSC_current_limit2020}. The outputs of the current controller are the $d-q$ set points of the modulated voltage, which are the inputs of the voltage modulation process of the \ac{GFVSC} ((c) in Fig.~\ref{fig:VSC_L_control_loops_general}).

Finally, the angle ($\theta$) used for Park's transformation from $d-q$ components to three-phase signal is provided by the \ac{GFVSC} self-synchronisation strategy ((d) in Fig.~\ref{fig:VSC_L_control_loops_general}), which will be described in the next section. Notice that the frequency imposed by the modulated voltage of the \ac{GFVSC}, $\omega$, is the angular speed of the mobile $d-q$ reference frame used for vector control and it is determined by the self-synchronisation strategy.  

\section{Self\hyp{}synchronisation control strategies of GFM\hyp{}VSCs}\label{sec:sscs_VSM}

\begin{figure}[htbp]
\centering
\includegraphics[width=0.7\textwidth]{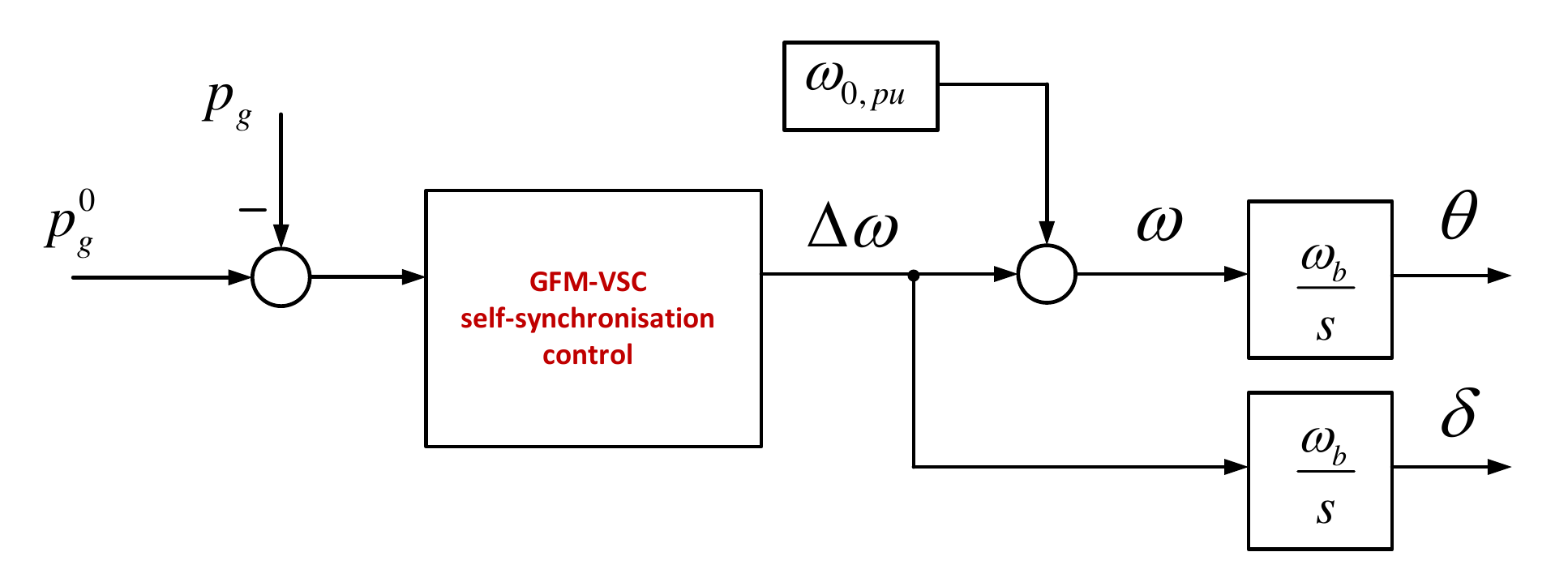}
\caption{General block diagram of a self-synchronisation control in \acp{GFVSC}.}
\label{fig:self-synchronisation_control}
\end{figure}

\noindent Fig.~\ref{fig:self-synchronisation_control} shows the general scheme of a self-synchronisation control, where the \ac{GFVSC} imposes its frequency according to a specific control algorithm, ensuring synchronism with the rest of the system but maintaining its behaviour as a voltage source behind a series impedance. The active-power setpoint ($p_{g}^{0}$) is compared to the measured active-power injection ($p_{g}$) and the error is the input of the self-synchronisation strategy. The outputs are the frequency ($\omega$, in pu) and angle ($\theta$, in rad) of the modulated voltage of the \ac{GFVSC}. The latter is used for Park's transformation (see Fig.~\ref{fig:VSC_L_control_loops_general}). The angle $\delta$ is the angle of the modulated voltage with respect to a synchronous reference frame.

The objectives of \ac{GFVSC} self-synchronisation strategies are as follows:
\begin{itemize}
    \item To produce synchronising effect and ensure that the \ac{GFVSC} reaches the same frequency than the rest of the system in steady state.
    \item When required, to \color{black} produce an inertial response of the \ac{GFVSC}.
    \item To control the dynamic response of the active- power injection \color{black} of the \ac{GFVSC} with sufficient damping, according to design specifications.
\end{itemize}
A primary frequency controller (or Primary Frequency Response, PFR) can also be included to the \ac{GFVSC}. Nevertheless, this controller is considered as a different part and it is not included in the self-synchronisation strategy itself. 

\begin{figure}[h!t]
\small
\centering
\begin{minipage}[b]{0.51315\linewidth}
  \includegraphics[width=\linewidth]{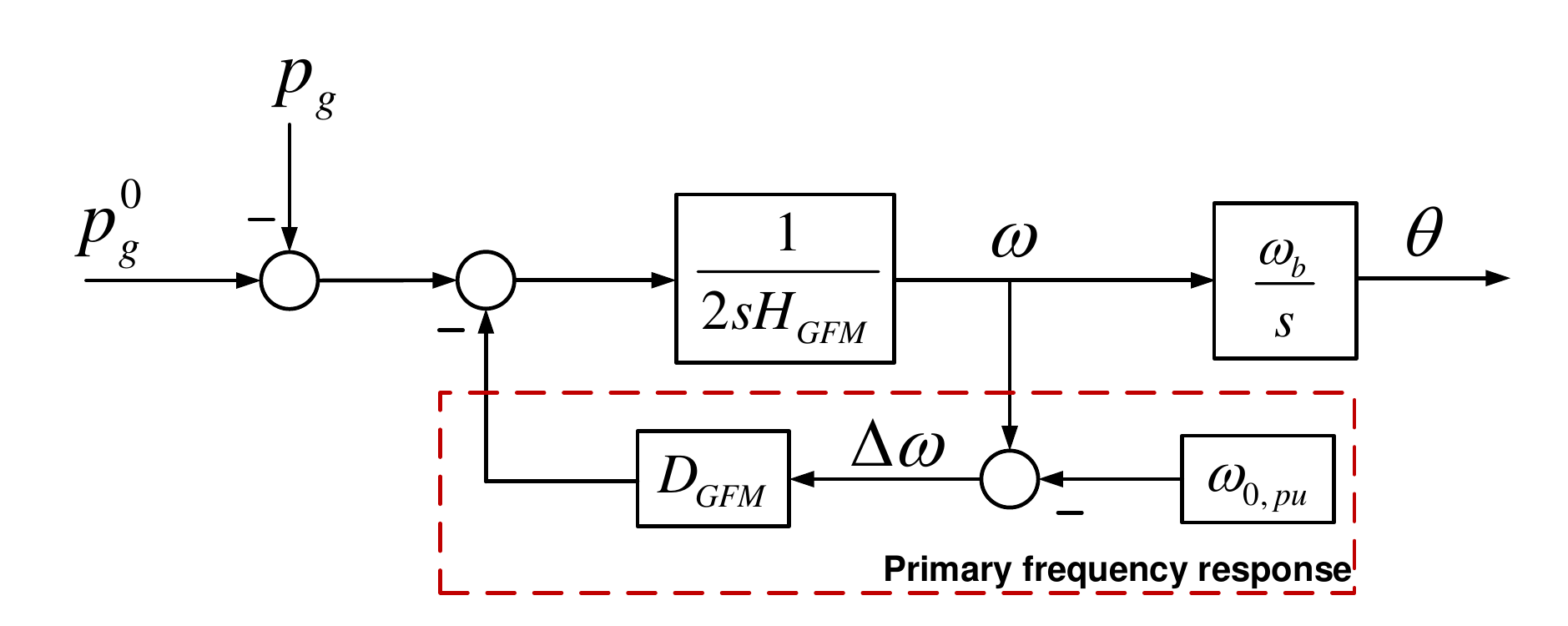}
  \caption*{(a) Strategy 1: VSM-noPLL.}
  \label{fig:imagen1}
\end{minipage}
\hfill
\begin{minipage}[b]{0.48\linewidth}
  \includegraphics[width=\linewidth]{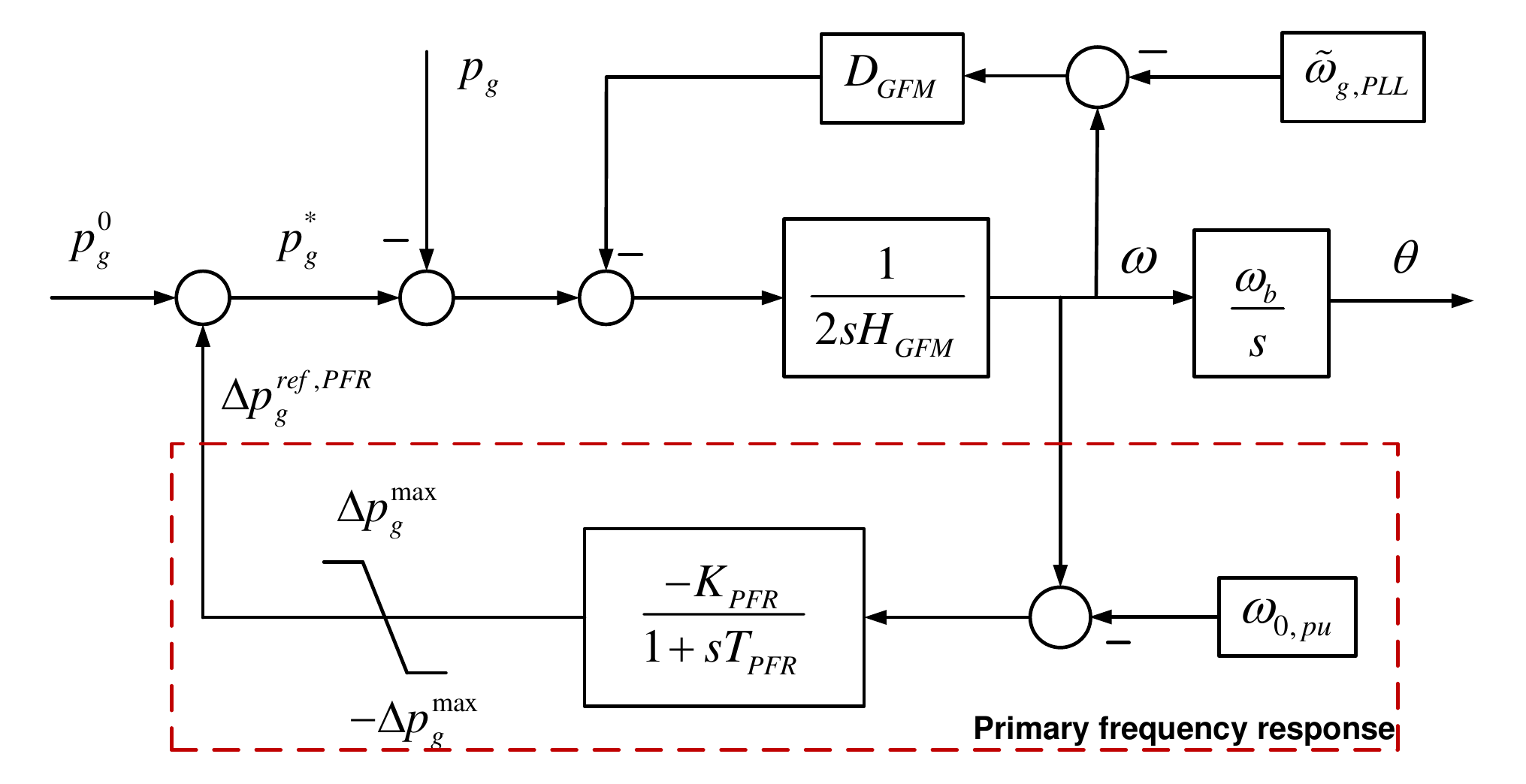}
  \caption*{(b) Strategy 2: VSM-PLL}
  \label{fig:imagen2}
\end{minipage}

\vspace{1em}

\begin{minipage}[b]{0.48\linewidth}
  \includegraphics[width=\linewidth]{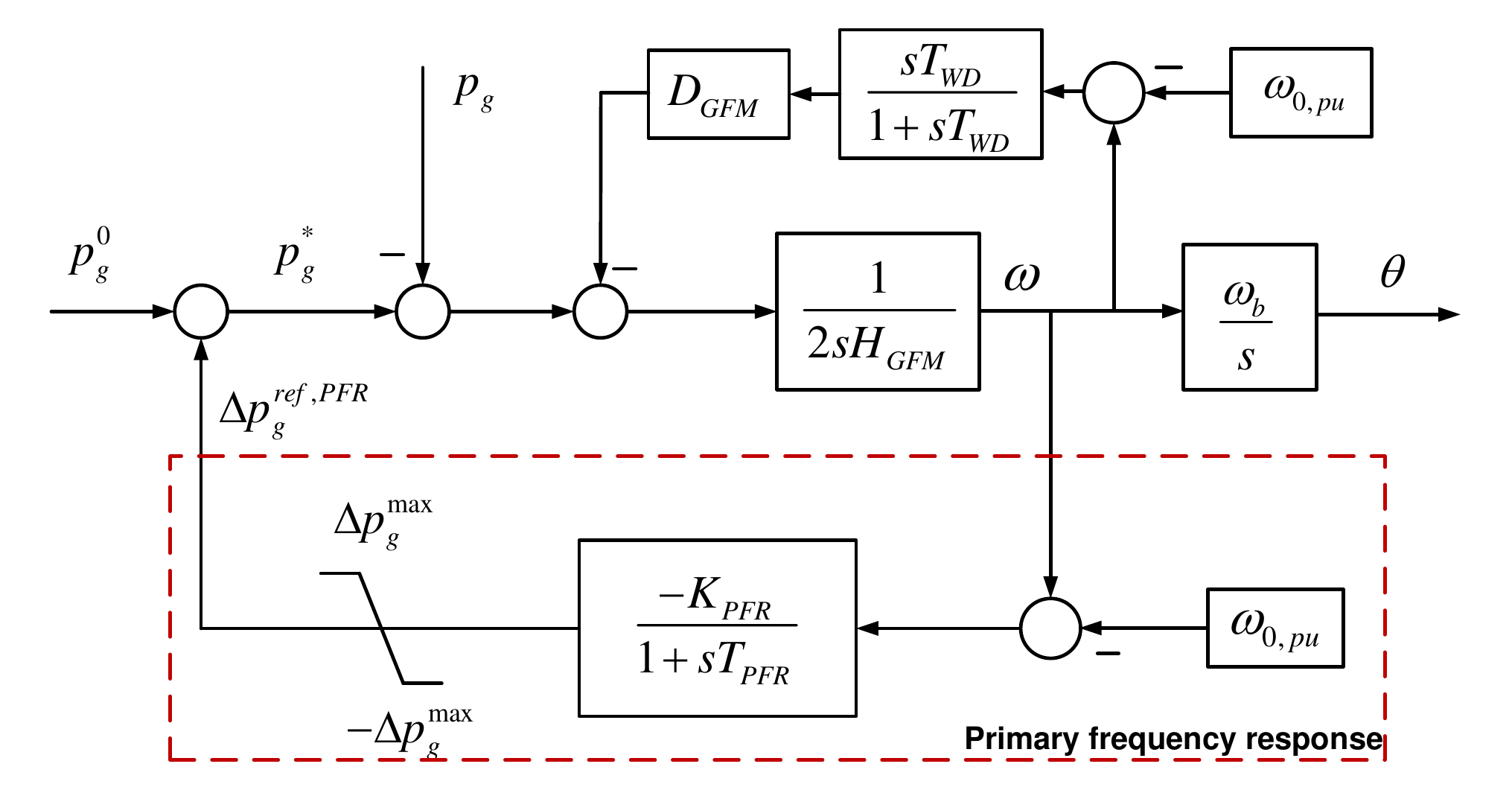}
  \caption*{(c) Strategy 3: VSM-Washout.}
  \label{fig:imagen3}
\end{minipage}
\hfill
\begin{minipage}[b]{0.48\linewidth}
  \includegraphics[width=\linewidth]{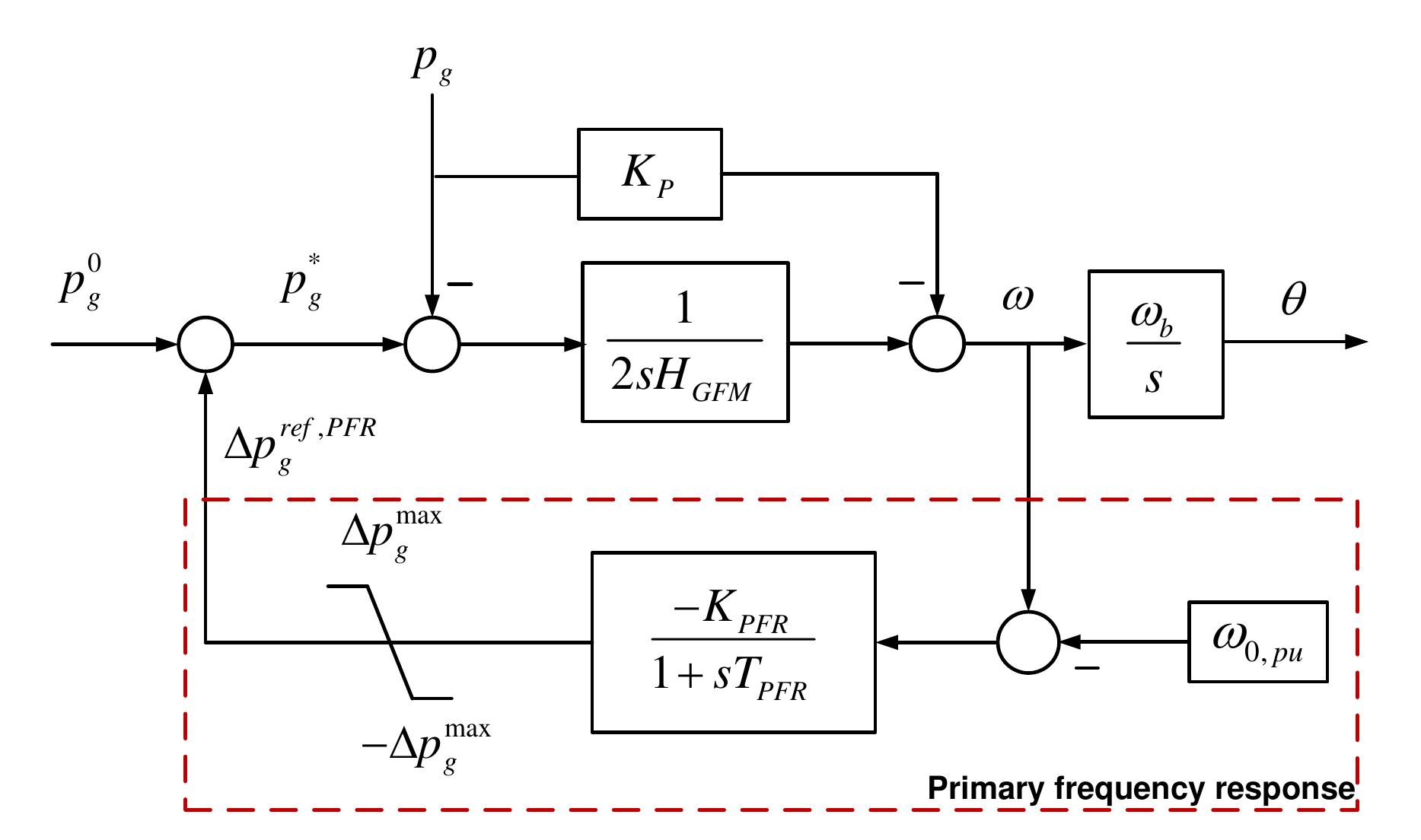}
  \caption*{(d) Strategy 4: IP control.}
  \label{fig:imagen4}
\end{minipage}

\caption{Different self-synchronisation control strategies in GFM-VSCs with primary frequency response (PFR) control.}
\label{fig:VSM_configurations}
\end{figure}

Some of the most extended self-synchronisation strategies for \acp{GFVSC} will be analysed in this work, namely:

\begin{itemize}
    \item \textbf{Strategy 1}: Virtual Synchronous Machine (VSM) without Phase-Locked Loop (PLL), hereinafter referred to as \textbf{VSM-noPLL} (see Fig.~\ref{fig:VSM_configurations}-(a)).
    \item \textbf{Strategy 2}: VSM with PLL, referred to as \textbf{VSM-PLL} (see Fig.~\ref{fig:VSM_configurations}-(b)).
    \item \textbf{Strategy 3}: VSM without PLL combined with a washout filter, referred to as \textbf{VSM-Washout} (see Fig.~\ref{fig:VSM_configurations}-(d)).
    \item \textbf{Strategy 4}: Integral Proportional (IP) controller, referred to as \textbf{IP control} (see Fig.~\ref{fig:VSM_configurations}-(c)).
\end{itemize}

Of the four self-synchronisation strategies, only strategy 1, VSM-noPLL (Fig.~\ref{fig:VSM_configurations}-(a)), which is equivalent to P-f droop control, has an implicit frequency control mechanism. This is because the inherent coupling between frequency dynamics and active power control allows the system to naturally respond to frequency deviations without the need for an additional control loop. In contrast, strategies 2, 3 and 4 (Fig.~\ref{fig:VSM_configurations}-(b)-(c)-(d)) decouple frequency and active power dynamics by incorporating a PLL (or another type of frequency estimator), a washout filter or a specific control structure, such as an IP controller. As a result, an external frequency control loop may be required to ensure adequate frequency regulation.
Therefore, an explicit frequency control loop consisting of a proportional gain ($K_{PFR}$), a low-pass filter with a time constant ($T_{PFR}$) and a saturation limit ($\Delta p_g^{max}$) is implemented in these cases. In Fig.~\ref{fig:VSM_configurations}, $\omega_b$ is the nominal frequency in rad/s.
\color{black}

The details of each \ac{GFVSC} \color{black} self-synchronisation strategy are provided in next subsections.

\FloatBarrier

\subsection{Strategy 1: VSM-noPLL }\label{sec:sscs_VSM_no_PLL}
Fig.~\ref{fig:VSM_configurations}-(a) shows the block diagram of the VSM without PLL. In this strategy, the \ac{GFVSC} mimics the behaviour of a synchronous generator and it imposes its frequency according to the swing equation \ac{GFVSC}~\cite{DArco2014, Choopani2020}. 

The main characteristics of VSM-noPLL are as follows:
\begin{itemize}
    \item The \ac{GFVSC} provides an inertial response, due to the emulated inertia constant $H_{GFM}$~(s).
    \item The damping factor coefficient, $D_{GFM}$, plays two roles simultaneously: (a) to provide damping and (b) to contribute to an instantaneous primary frequency response. Notice that it is mandatory that the primary frequency support is instantaneous, because, otherwise, the dynamic response of the \ac{GFVSC} could be jeopardised. Notice also that if a typical value for the proportional gain of the frequency controller is used (e.g. $D_{GFM}=20$~pu), then the dynamic response of the \ac{GFVSC} is determined. 
\end{itemize}

\FloatBarrier
\vspace{-0.2cm}

\subsection{Strategy 2: VSM-PLL}\label{sec:sscs_VSM_PLL}
\begin{figure}[!htbp]
\begin{center}
\includegraphics[width=0.75\columnwidth]{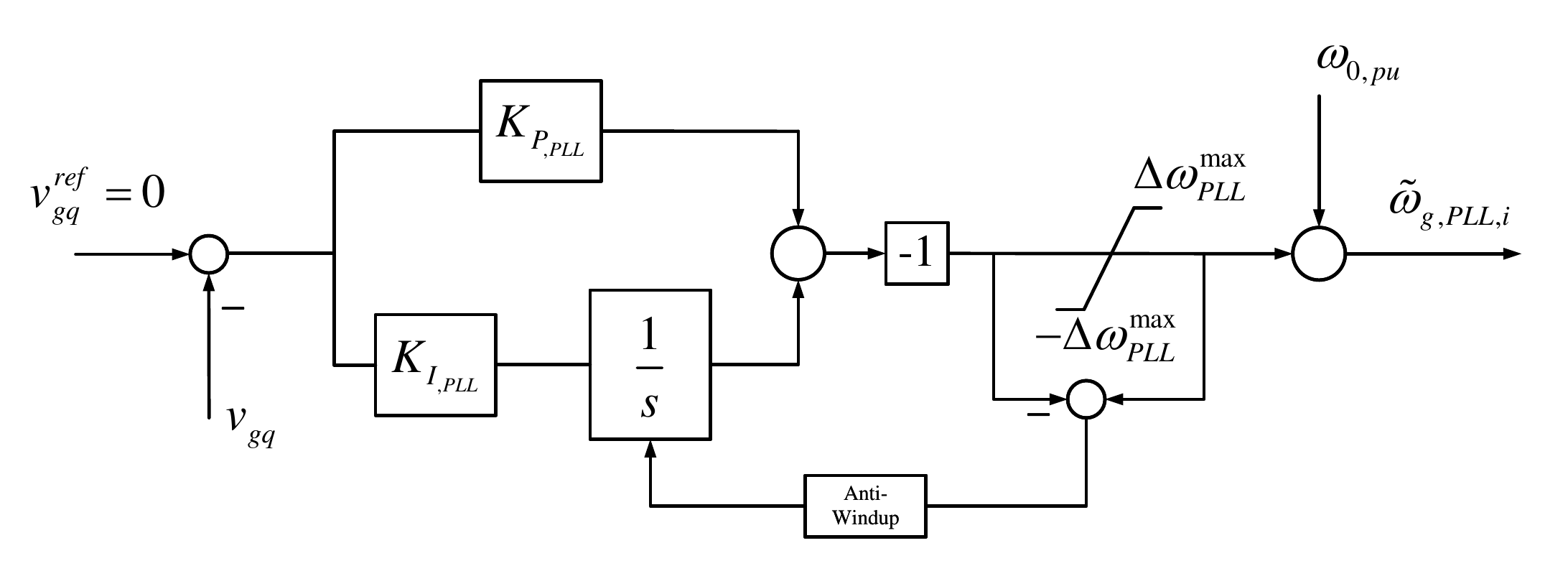}
\caption{PLL block diagram equipped with an anti-windup.}
\label{fig:PLL_for_VSM}
\end{center}
\end{figure}

Fig.~\ref{fig:VSM_configurations}-(b) shows the block diagram of the VSM with PLL, equipped with PFR control. In this strategy, the \ac{GFVSC} imposes its frequency according to the swing equation, but using a PLL to estimate the frequency at the PCC~\cite{DArco2015, Rokrok2020}. Then, the difference between the frequency of the \ac{GFVSC} and the estimated frequency of the grid at the PCC is multiplied by the damping coefficient $D_{GFM}$. The block diagram of the PLL equipped, which includes an anti-windup method, is shown in Fig.~\ref{fig:PLL_for_VSM}. Notice that, although this \ac{GFVSC} self-synchronisation scheme uses a PLL, the application is totally different to the use of a PLL in grid-following converters. Here is only used to estimate the frequency of the grid and to be used in the self-synchronisation strategy.

The main characteristics of VSM-PLL are as follows
\begin{itemize}
    \item The \ac{GFVSC} needs an estimation of the frequency at the AC connection point, using a PLL.  \color{black}
    \item The \ac{GFVSC} provides an inertial response, due to the emulated inertia constant $H_{GFM}$~(s).
    \item The damping coefficient, $D_{GFM}$, is used to obtain the required damping ratio of the dynamic response of the \ac{GFVSC}.
    \item The controller of the primary frequency response of the \ac{GFVSC} is decoupled from the power control. For example, the PFR control can be made as slow as required (time constant $T_{PFR}$), without jeopardising the synchronising properties of the strategy. Notice also, that if primary frequency controller is not used (e.g. $K_{PFR}=0$~pu), the \ac{GFVSC} controls its active-power injection to a constant value (e.g. $p_{g}=p_{g}^0$).
\end{itemize}

\FloatBarrier

\FloatBarrier

\subsection{Strategy 3: VSM-Washout
}\label{sec:sscs_VSM_no_PLL_TDM}
Fig.~\ref{fig:VSM_configurations}-(c) shows the scheme of the VSM without PLL, but incorporating a wash-out filter with a time constant $T_{WD}$ \cite{MoO_DArco2017,DIgSILENT2022}. This high-pass filter remains inactive during steady states but it acts during the transient. 

The main characteristics of VSM-Washout are as follows:
\begin{itemize}
    \item The \ac{GFVSC} does not use a PLL.
    \item The \ac{GFVSC} provides an inertial response, due to the emulated inertia constant $H_{GFM}$~(s).
    \item The damping coefficient, $D_{GFM}$, is used to obtain the required damping ratio of the dynamic response of the \ac{GFVSC}.
    \item The primary frequency controller of the \ac{GFVSC} is decoupled from the power control, due to the washout filter. 
\end{itemize}

\FloatBarrier

\subsection{Strategy 4: IP control}\label{sec:sscs_IP_control}
The IP controller is another approach for controlling \ac{GFVSC} power converters described in~\cite{Qoria2020}. This type of controller adds a proportional action ($K_P$) to the active power to provide damping, as shown in Fig.~\ref{fig:VSM_configurations}-(d). 

\FloatBarrier
The main characteristics of IP control are as follows:
\begin{itemize}
    \item The \ac{GFVSC} does not require an estimation of the frequency at the AC connection point. \color{black}
    \item The \ac{GFVSC} provides an inertial response, due to the emulated inertia constant $H_{GFM}$~(s).
    \item The proportional gain, $K_P$, is used to obtain the required damping ratio of the dynamic response of the \ac{GFVSC}.
    \item The primary frequency controller of the \ac{GFVSC} is decoupled from the self-synchronisation control. Then an external control loop for PFR support can be added to the GFM-VSC control (Fig.~\ref{fig:VSM_configurations}~(c)).
    \item As will be explained in Section~\ref{sec:sscs_equivalences}, the behaviour of IP controller under small disturbances is equivalent to the behaviour of VSM-PLL strategy \cite{Qoria2020}.
\end{itemize}


\subsection{Design of self-synchronisation strategies}\label{sec:sscs_equivalences}
\noindent Parameters of the \ac{GFVSC} self-synchronisation \color{black} strategy will be designed as follows:
\begin{itemize}
    \item To obtain a dynamic response of the \ac{GFVSC} with the required characteristics.
    \item When possible, the PFR controller is designed separately, using a typical values for the proportional gain.
\end{itemize}
The analysis is based on the one carried out in previous work \cite{Qoria2020,Strunz2023} and the details are omitted in this paper.

The system of Fig.~\ref{fig:VSC_V_model_sscs} is considered. Firstly, a \ac{GFVSC} with VSM-PLL is considered. The PFR controller is assumed to be slow and it is neglected from the analysis. It can be shown that, under small disturbances, the transfer function between the active-power set point and output of the \ac{GFVSC} ($T_{(s)}=p_{g}/p_{g}^*$) can be written as \cite{Qoria2020,Strunz2023}:
\begin{equation}
T (s) = \frac{p_{g} (s)}{p_{g}^* (s)} =  \scalebox{1.5}{$\frac{\left(\frac{\omega_b}{2H_{GFM}x_{c}}\right)}{s^2 + \frac{D_{GFM}}{2H_{GFM}}s + \frac{\omega_b}{2H_{GFM}x_{c}}}$} \label{eq:transfer_function}
\end{equation}
Hence, the damping ratio, $\zeta$, and the natural frequency, $\omega_n$, are given by:
\begin{equation}
\zeta = \frac{D_{GFM}}{4H_{GFM}}
\frac{1}{\omega_n}, \mbox{\space \space}\omega_n = \sqrt{\frac{\omega_b}{2H_{GFM}x_{c}}}. \label{eq:VSM_zeta_wn}
\end{equation}
There are two equations (\ref{eq:VSM_zeta_wn}) and four parameters ($H_{GFM}$, $D_{GFM}$, $\zeta$ and $\omega_n$).

In the case of VSM-noPLL strategy, under the assumption that the estimated frequency at the PCC is similar to the frequency of the infinite grid, which is an approximation assuming a strong grid, Eq.~(\ref{eq:VSM_zeta_wn}) also holds. In the case of VSM-Washout strategy, the same approximation is made plus the approximation that the wash-out filter has low effect and it is neglected. This is only true if the wash-out time constant ($T_{WD}$) is high enough. \color{black} Then, Eq.~(\ref{eq:VSM_zeta_wn}) can also be used. Further considerations about the design aspects and the wash-out-filter time constant in VSM-Washout are described in Subsection~\ref{sec:sscs_gfm_vsc_design_VSM_washout}. \color{black}

The work in \cite{Qoria2020} proved the equivalence between VSM-PLL and IP control, with where $D_{GFM} = 2H_{GFM}K_{P}\omega_b/x_{c}$. Therefore, in IP control the damping ratio and natural frequency are related to the control parameters as follows \cite{Qoria2020}:
\begin{equation}
K_{P} = \zeta\sqrt{\frac{2x_{c}}{H_{GFM}\omega_b}} \quad \text{and} \quad \omega_n = \sqrt{\frac{\omega_b}{2H_{GFM}x_{c}}}. \label{eq:IP_controller_zeta_wn}
\end{equation}

Therefore, the design of GFM-VSC self-synchronisation strategies used in this work is as summarised in Table \ref{tab:GFM_VSC_self_sync_design}.

\begin{table}[htbp]
    \caption{Design of GFM-VSC self-synchronisation strategies.} 
    \begin{center}
    \scalebox{0.85}{
    \begin{tabular}{|l|c|c|}
    \hline
    \multirow{2}{*}{\textbf{Strategy}} &  Specified & Calculated \\ 
    & values & values  \\ 
    \hline
    \multirow{4}{*}{Strategy 1: VSM-noPLL} & $H_{GFM}$, $D_{GFM}$& \multirow{4}{*}{$\zeta$ and $\omega_n$, with (\ref{eq:VSM_zeta_wn})}  \\
     & PFR (implicit):  &  \\
     & $K_{PFR}=D_{GFM}$&   \\ 
     & $T_{PFR}=0$ s &   \\  \hline
    \multirow{2}{*}{Strategy 2: VSM-PLL} & $H_{GFM}$, $\zeta$ & \multirow{2}{*}{$D_{GFM}$ and $\omega_n$, with with (\ref{eq:VSM_zeta_wn})}  \\
     & PFR: $K_{PFR}$, $T_{PFR}$  &   \\ \hline
     \multirow{3}{*}{Strategy 3: VSM-Washout} & $H_{GFM}$, $\zeta$ & \multirow{3}{*}{$D_{GFM}$ and $\omega_n$, with (\ref{eq:VSM_zeta_wn})}  \\
    & PFR: $K_{PFR}$, $T_{PFR}$ &   \\ 
     & Wash-out: $T_{WD}$ &   \\
    \hline
    \multirow{2}{*}{Strategy 4: IP control} & $H_{GFM}$, $\zeta$& \multirow{2}{*}{$K_P$ and $\omega_n$, with (\ref{eq:IP_controller_zeta_wn})}  \\
     & PFR: $K_{PFR}$, $T_{PFR}$ &   \\ \hline
    \end{tabular}
    }
    \label{tab:GFM_VSC_self_sync_design}
    \end{center}
    \end{table}

\subsection{Design of self-synchronisation strategies - Specific aspects of Strategy 3 (VSM-Washout)}\label{sec:sscs_gfm_vsc_design_VSM_washout}
\noindent As described previously, Strategy 3 (VSM-Washout), Eq.~(\ref{eq:VSM_zeta_wn}) and Table~\ref{tab:GFM_VSC_self_sync_design} could also be used for the design, taking into account small-signal stability aspects and using the assumption that the wash-out-filter ($T_{WD}$) is high enough. Nevertheless,  \color{black} in Strategy 3 (VSM-Washout), transient-stability behaviour should be also taken into account. In previous work, the design of the wash-out filter time constant in strategy VSM-Washout has not been addressed explicitly. However, reference~\cite{Xiong2021} has analysed the impact of the wash-out filter in the so-called transient damping method (TDM), and conclusions can be applied directly to the VSM-Washout strategy. The work in~\cite{Xiong2021} proposes a curve for feasible regions of parameters $D_{GFM}$ and $T_{WD}$, in terms of transient stability. Mainly,in order to use high values of $D_{GFM}$, the wash-out filter time constant should be high enough, concluding that $T_{WD}\ge 1$ s could be a reasonable value, for the test system analysed. In addition, in Strategy 3 the wash-out filter adds a state variable and the closed-loop system dynamics is more complex and it is not a second-order system any more. However, this aspect can be addressed by using a high enough value of $T_{WD}$ and it is recommended to check the results of the design by time-domain simulation. Obviously, the wash-out filter time constant $T_{WD}$ should not be excessively high, because it would not be able to avoid active-power changes under frequency offsets, which is its role. \color{black}

\FloatBarrier
\newpage
\section{Methods to improve transient stability}\label{sec:sscs_VAPC_FL}
\noindent This paper also analyses two methods to improve transient stability in \acp{GFVSC}, which could be applied to all \ac{GFVSC} self-synchronisation strategies of Section~\ref{sec:sscs_VSM}:
\begin{itemize}
    \item Virtual active power control (VAPC), proposed in previous work~\cite{VattaKkuniK2024}.
    \item A frequency limiter for \ac{GFVSC} control (FLC), \color{black} proposed in this paper.
\end{itemize}
\subsection{Virtual active power control (VAPC)}\label{sec:sscs_VAPC_Controller}
\begin{figure}[htbp]
\centering
\includegraphics[width=0.7\textwidth]{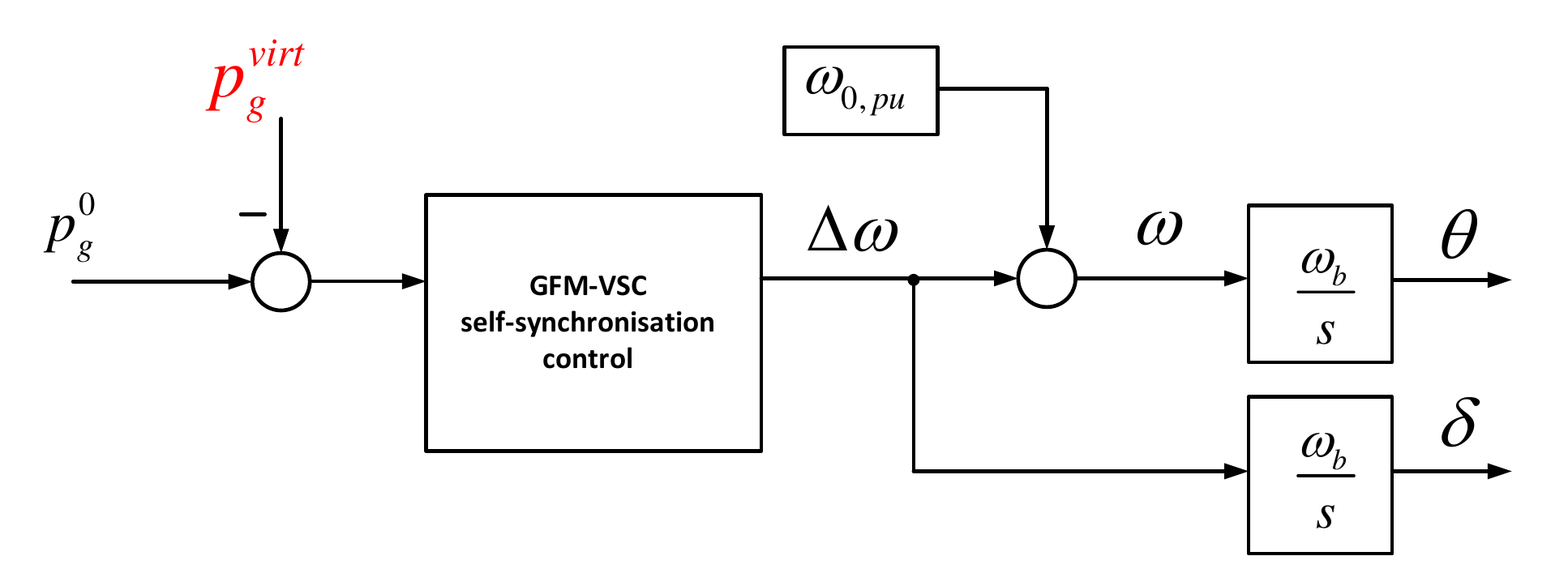}
\caption{General block diagram of a self-synchronisation control in \acp{GFVSC} using the virtual active power $p_{g}^{virt}$ as a feedback measurement.}
\label{fig:self-synchronisation_control_Pvirt_sscs}
\end{figure}

\noindent The virtual active power control (VAPC) was proposed in~\cite{VattaKkuniK2024} to improve transient stability in \acp{GFVSC}. In VAPC, the \ac{GFVSC} uses as feedback signal of the self-synchronisation mechanism the \emph{unsaturated virtual active power}, $p_{g}^{virt}$, instead of the measured active power injection,  as shown in Fig.~\ref{fig:self-synchronisation_control_Pvirt_sscs}. Notice that VAPC is compatible with any self-synchronisation mechanism in \acp{GFVSC}.  

The unsaturated virtual active power used in VAPC ($p_{g}^{virt}$ in Fig.~\ref{fig:self-synchronisation_control_Pvirt_sscs}) is calculated as follows:
\begin{equation}\label{eq:Pv}
p_{g}^{virt} = v_{g,d} \cdot i_{g,d}^{ref'}+v_{g,q} \cdot i_{g,q}^{ref'}
\end{equation}
where $v_{g,d}$ and $v_{g,q}$  are the voltage components at PCC and $i_{g,d}^{ref'}$ and $i_{g,q}^{ref'}$ are the unsaturated current set points. 

All details and analysis of VAPC control can be found in~\cite{VattaKkuniK2024}. 

\subsection{Proposed frequency limiter control (FLC)}\label{sec:sscs_freq_lim}
\noindent In this Section, a frequency limiter (FLC) \color{black} in \acp{GFVSC} is proposed to improve transient stability. The proposed frequency limiter was motivated by previous work in~\cite{Qoria2020,Collados-Rodriguez_2023,Du2019,Du2024,XZhao_2020,XZhao_2022,Saman_2022,Li2024}, but it introduces new features to improve its performance, as discussed in Section~\ref{sec:sscs_intro}. The proposed frequency limiter in \acp{GFVSC}: (a) is aimed to improve transient stability, (b) it is activated during when a fault is detected, by looking to the terminal voltage and using a hysteresis characteristic, and (c) it includes a frequency saturator around the pre-fault steady-state frequency. 

\begin{figure}[htbp]
\centering
\includegraphics[width=0.65\textwidth]{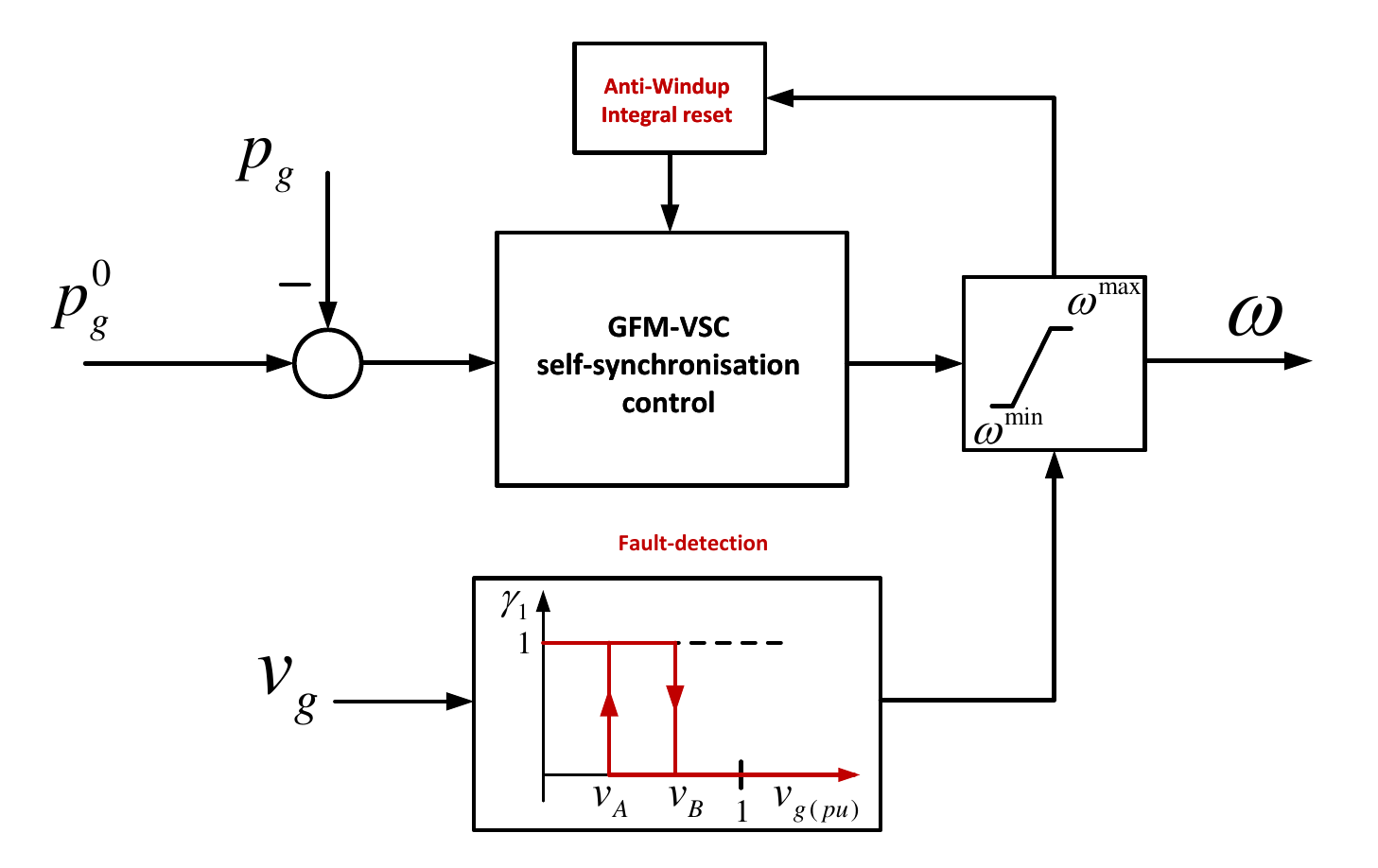}
\caption{Proposed frequency limiter applied in GFM-VSC self-synchronisation strategies.}
\label{fig:frequencylimiter_sscs}
\end{figure}

Fig.~\ref{fig:frequencylimiter_sscs} shows the block diagram of the proposed algorithm for frequency limitation in \acp{GFVSC} to improve transient stability. The proposed frequency limiter consists of the following elements:
\begin{itemize}
    \item Fault Detection: It is performed with a hysteresis characteristic:
    \begin{itemize}
        \item A binary variable $\gamma_{1}$ is used, which is set to 1 upon detection of a voltage decrease with a hysteresis, as illustrated in Fig.~\ref{fig:frequencylimiter_sscs}.
        \item If $v_{g} \leq v_{A}$ (fault detected), then $\gamma_{1}=1$ and remains at 1 until $v_{g}>v_{B}$. If a fault is not detected, $\gamma_{1}=0$.
    \end{itemize}
    \item Frequency Saturator: It is activated only when the fault is detected, limiting the frequency if it exceeds a specific threshold: $\omega^{max}, \omega^{min}$.
    \item Anti-Windup mechanism: When the frequency saturates, it resets the integral part of the self-synchronisation mechanism (i.e., the integral of the virtual inertia).
\end{itemize}
The frequency limits of the proposed current-limitation algorithm is given by:
\begin{eqnarray}\label{eq_omega_max_min}
\begin{array}{cc}
\omega^{max} = & \omega^{steady-state} + \Delta \omega^{max}\\
\\
\omega^{min} = & \omega^{steady-state} - \Delta \omega^{max}
\end{array}
\end{eqnarray}
where:
\begin{itemize}
    \item \(\omega^{max}\) is the maximum saturated frequency.
    \item \(\omega^{min}\) is the minimum saturated frequency.
    \item \(\omega^{steady-state}\) is the unsaturated frequency of the converter in steady-state before the fault. Notice that the steady-state pre-fault frequency is used here, to avoid dependency on steady-state frequency deviations in the system.
    \item \(\Delta \omega^{max}\) is the maximum increment of the frequency of the GFM-VSC. This parameter should be designed (e.g. $\Delta \omega^{max}=0.005$ pu, for example).
\end{itemize}

\FloatBarrier
\newpage

\section{Theoretical analysis}\label{sec:Theoretical_analysis}
\noindent This section presents a theoretical analysis on the impact on transient stability of self-synchronisation strategies in \acp{GFVSC}, the concept of virtual active power control (VAPC) and the proposed frequency limiter for \acp{GFVSC}. Active power - angle ($P-\delta$) curves are used for the analysis, following the guidelines used in previous work~\cite{Qoria_VSC_current_limit2020,Qoria_VSC_CCT2020,VattaKkuniK2024,XiongX_2021a,BlaabjergFTSAngle2022, SiW2023,RAvilaM2024}. In synchronous machines, Equal Area Criterion (EAC) is used for the theoretical transient stability analysis~\cite{Kundur1994,Exposito2018}. However, although the EAC could provide some qualitative insight of transient-stability behaviour of \ac{GFVSC} power converters, it cannot be applied directly to \acp{GFVSC}, as shown in~\cite{Chatterjee2025}. This is mainly due to the fact that the high damping coefficient used in \acp{GFVSC} is much higher than the one in synchronous machines. This implies that, after the fault clearing, the angle of the \ac{GFVSC} increases much less than the rotor angle in synchronous machines, as illustrated in~\cite{Carmen_Cardozo2024}. 

Hence, the key aspects taken into account for the theoretical analysis are as follows:
\begin{itemize}
    \item $P-\delta$ curves: The larger is the $P-\delta$ curves, the larger is transient-stability margin of the \ac{GFVSC} and this will result in a higher critical clearing time (CCT) of the fault, as shown in previous work~\cite{Qoria_VSC_current_limit2020,Qoria_VSC_CCT2020,VattaKkuniK2024,XiongX_2021a,BlaabjergFTSAngle2022, SiW2023,RAvilaM2024}.
    \item Structure and differential equation of the \ac{GFVSC} self-synchronisation strategy and methods to improve transient stability: if a fault occurs, the best results in terms of transient stability are obtained with strategies in which the frequency and angle increase less, during the fault and after the fault clearing. This depends on each stategy.
\end{itemize}
\color{black}

\begin{figure}[!htbp]
	\begin{center}
		\centering
		\includegraphics[width=0.7\columnwidth]{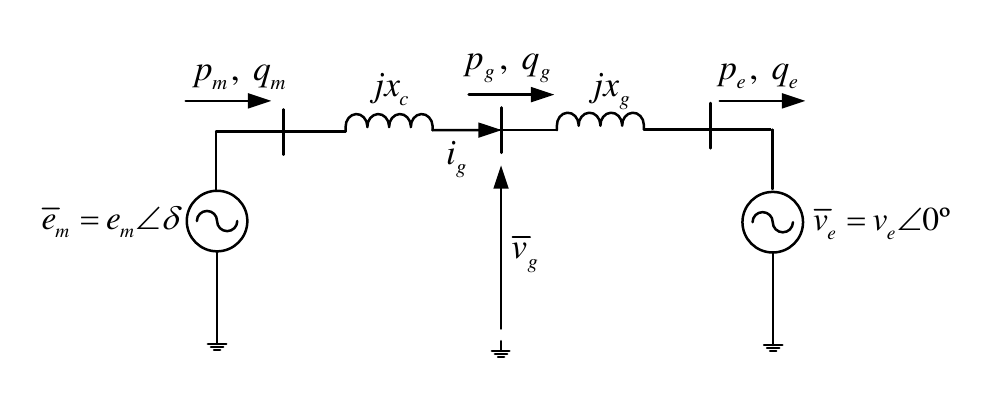}
		\caption{GFM-VSC connected to an infinite grid.}
		\label{fig:gfm_vsc_smib_sld}
	\end{center}
\end{figure}

A \ac{GFVSC} is connected to an infinite grid, as shown in Fig.~\ref{fig:gfm_vsc_smib_sld}. It is assumed that the external grid is strong ($x_g$ low) and the series resistances are neglected, for illustrative purposes. The control system of the \ac{GFVSC} is as the one described in Fig.~\ref{fig:VSC_L_control_loops_general} of Section~\ref{sec:VSC_V}, where the virtual impedance of the quasi-static model is assumed to be zero ($x_v=0$), hence $x_{cv}=x_c$. A conventional Current Saturation Algorithm (CSA) is used for current limitation of the \ac{GFVSC}, with equal priority to $d$- and $q$-axis current injections. The magnitude of modulated voltage of the \ac{GFVSC} is assumed to be constant for the analysis ($e_{m}=e_{m}^{0}$).

The active power injection of the \ac{GFVSC}, when the current limit has not been reached, reads:
\begin{equation}\label{eq:VSC_pg_EAC}
	p_{g} = p_{e}= \frac{e_{m} v_{e}}{x_{tot}} \sin \delta
\end{equation}
where $\delta$ is the angle of the modulated voltage of the \ac{GFVSC} (with respect to the external grid) and $x_{tot}=x_{c}+ x_{g}\color{black}$.

A three-phase-to-ground short circuit is applied at bus $e$  at a certain time $t_{f}$ (for illustrative purposes) and it is cleared at a certain time $t_{cl}$ . When the fault is cleared, the angle of the modulated voltage of the \ac{GFVSC} is called $\delta_{cl}$. When the fault occurs, $v_e=0$ and, therefore, the active-power injection of the \ac{GFVSC} goes to zero too: $p_g=0$. 

When the \ac{GFVSC} has current limiter CL-CSA, the active-power injection of the \ac{GFVSC} is given by~\cite{VattaKkuniK2024,LabaY2023}:
\begin{equation}\label{eq:GFM_pg_case2}
\begin{split}
    p_{g} = \left\lbrace\begin{array}{ccc} 
        \frac{e_{m}^{0} v_{e}}{x_{\text{tot}}} \sin \delta, &\text{if}& \delta < \delta_a \text{ } (i_g < i_g^{\max}) \\\\
        \frac{e_{m}^{0} v_{e} \sin \delta}{\sqrt{(e_{m}^{0})^2 + v_{e}^2-2e_{m}^{0} v_{e} \cos \delta}} \cdot i_{g}^{\max}, &\text{if}& \delta \geq \delta_a \text{ } (i_g = i_g^{\max})
    \end{array}\right.
\end{split}
\end{equation}
where $\delta_a$ is the angle at which the current injection of the \ac{GFVSC} reaches its current limit ($i_g = i_g^{\max}$). 

\FloatBarrier
\begin{figure}[!htbp]
	\begin{center}
		\centering\includegraphics[width=0.8\columnwidth]{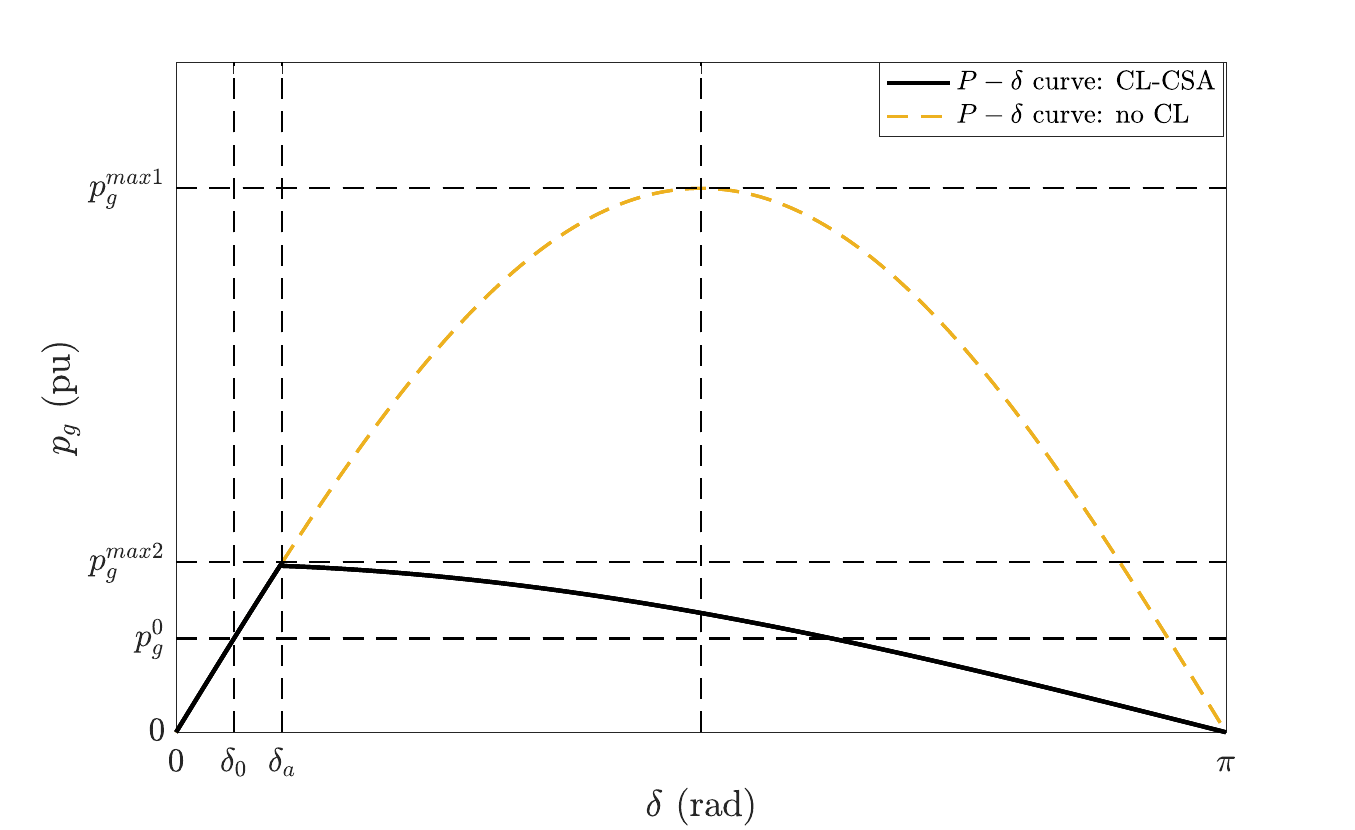}
		\caption{\ac{GFVSC} with CL-CSA: $P-\delta$ curve.}
		\label{fig:EAC_case2}
	\end{center}
\end{figure}

\FloatBarrier
Fig.~\ref{fig:EAC_case2} shows the the $P-\delta$ curve of the \ac{GFVSC} with current limiter CL-CSA. The $P-\delta$ curve of the \ac{GFVSC} without current limiter is also represented (dashed line in yellow), for comparison purposes. In the initial operating point: $\delta=\delta_0$ and $p_g=p_g^0$. Angle $\delta_{cl}$ is defined as the angle when the fault is cleared. Without CL, the $P-\delta$ curve of the \ac{GFVSC} is a sinusoidal curve, given by (\ref{eq:VSC_pg_EAC}), where the maximum active-power injection is $p_{g}^{max1}=e_{m}^{0} v_{e}/x_{tot}$. When the \ac{GFVSC} has CL-CSA, the are of the $P-\delta$ curve is reduced dramatically and the maximum active-power injection of the \ac{GFVSC} is given by  $p_{g}^{max2}=e_{m}^{0} v_{e} \sin(\delta_a)/x_{tot}<p_{g}^{max1}$.

The larger is the $P-\delta$, the larger will be the critical angle $\delta_{cl}^{crit}$ (the maximum angle that can take  place before clearing the short circuit, without provoking loss of synchronism of the \ac{GFVSC}) and the larger will be CCT. It is well-known that when the \ac{GFVSC} is equipped with current limiter CL-CSA, its transient-stability margin is reduced, as can be observed in Fig.~\ref{fig:EAC_case2} (see~\cite{VattaKkuniK2024} for further details).   \color{black}

However, the CCT depends on the $P-\delta$ curve but also depends on the particular self-synchronisation strategy used in \ac{GFVSC} control (Fig.~\ref{fig:self-synchronisation_control} of Section~\ref{sec:sscs_VSM}). Specifically, the CCT can be obtained (sometimes analytically and sometimes numerically) using the swing equation of the \ac{GFVSC} self-synchronisation strategy and imposing the condition that the frequency of the \ac{GFVSC} can be recovered to its pre-fault value, after the fault clearing~\cite{Qoria_VSC_CCT2020}. 

\subsection{GFM-VSC self-synchronisation strategies}\label{sec:Theo_GFM_VSC_control}
A qualitative analysis of \ac{GFVSC} self-synchronisation strategies of Section~\ref{sec:sscs_VSM} (see Fig.~\ref{fig:VSM_configurations}) is presented in this subsection. For the analysis, it is assumed that the frequency controller of \ac{GFVSC} is slow and it has low impact on transient stability; therefore, it is neglected for simplicfication purposes. Naturally, in the only case that the frequency controller is considered is in the case of VSM without PLL, because it this case the damping term plays also the role of an instantaneous primary frequency response (PFR) term and it cannot be neglected. The evolution of the frequency and angle of the modulated voltage of the \ac{GFVSC} will be analysed in a qualitative way, and the CCT of each self-synchronisation strategy will be discussed. Notice that an analytical formula of the critical angle and the CCT cannot be obtained when considering the $P-\delta$ characteristic of the \ac{GFVSC} with CL-CSA (\ref{eq:GFM_pg_case2}) and the different differential equations of the self-synchronisation strategies (swing equations). However, useful insight can be obtained by means of qualitative analysis and numerical simulation.

\subsection{Strategy 1: VSM-noPLL }\label{sec:Theo_GFM_VSC_VSM_no_PLL}
In the VSM without PLL scheme (Fig.~\ref{fig:VSM_configurations}-(a)), when a fault occurs, the active-power injection of the \ac{GFVSC} goes to zero and the differential equation (swing equation) of this self-synchronisation strategy reads:
\begin{eqnarray}\label{eq:VSM_no_PLL_swing_eq}
	2 H_{GFM}\frac{d \omega}{dt} + D_{GFM} (\omega-\omega_{0,pu}) & = & p_{g}^{0} - \cancelto{0}{p_{g}}  
\end{eqnarray}
Therefore, the derivative of the frequency of the \ac{GFVSC} during the fault is given by:
\begin{eqnarray}\label{eq:VSM_no_PLL_swing_eq_v1}
    \frac{d \omega}{dt} = \frac{1}{2 H_{GFM}} [ p_{g}^{0} - D_{GFM} (\omega-\omega_{0,pu})  ]	
\end{eqnarray}
Clearly, high values of the emulated inertia constant ($H_{GFM}$) and the damping coefficient ($D_{GFM}$) prevent loss of synchronism, because they reduce acceleration of the frequency of the \ac{GFVSC} and, therefore, the increase of the angle of the modulated voltage of the \ac{GFVSC} ($\delta$) is also reduced. As a consequence, high values of $H_{GFM}$ and $D_{GFM}$ increase the critical clearing angle ($\delta_{cl}^{crit}$) and the CCT. However, this strategy has a limitation: since the damping coefficient $D_{GFM}$ plays two roles at the same time: the damping term and the instantaneous primary frequency controller, typical values for primary frequency control gains should be used ($D_{GFM}=20-25$~pu), and not higher. This aspects limits the performance of the \ac{GFVSC} in terms of transient stability.

\subsection{Strategy 2: VSM-PLL }\label{sec:Theo_GFM_VSC_VSM_with_PLL}
In the VSM with PLL scheme (Fig.~\ref{fig:VSM_configurations}-(b)), when a fault occurs, the swing equation of the \ac{GFVSC} is given by:
\begin{eqnarray}\label{eq:VSM_with_PLL_swing_eq}
	2 H_{GFM}\frac{d \omega}{dt} + D_{GFM} (\omega-\tilde{\omega}_{g,PLL}) & = & p_{g}^{0} - \cancelto{0}{p_{g}}  
\end{eqnarray}
Assuming that the grid is strong ($x_g$ low), the estimated frequency at the connection point by the PLL is similar to the nominal frequency: $\tilde{\omega}_{g,PLL}\approx \omega_{0,pu}$. Therefore, the behaviour of the VSM-PLL strategy is the same as that of the VSM scheme without PLL (given by equations (\ref{eq:VSM_no_PLL_swing_eq}) and (\ref{eq:VSM_no_PLL_swing_eq_v1})). However, there is an important difference, in the VSM-PLL scheme, the damping term (with damping coefficient $D_{GFM}$) is decoupled from the primary frequency controller. Hence, the damping coefficient could be much higher (e.g., $D_{GFM}=100-300$~pu). This prevents loss of synchronism, since, during the fault, the increase of the frequency and the angle of the \ac{GFVSC} is much lower is much lower than then one obtained with the VSM-noPLL strategy. Thus, the CCT achieved with the VSM-PLL strategy is expected to be higher than then one obtained with the VSM-noPLL strategy.

\subsection{Strategy 3: VSM-Washout }\label{sec:Theo_GFM_VSC_VSM_no_PLL_washout}
In the VSM without PLL + wash-out filter scheme (Fig.~\ref{fig:VSM_configurations}-(c)), the damping term and the frequency controller are also decoupled. Hence, a high value of the damping coefficient can be used ($D_{GFM}=100-300$~pu). For the analysis, the wash-out filter is neglected: $sT_{WD}/(1+sT_{WD}) \approx 1$. Therefore, the CCT obtained when using strategy VSM-Washout is expected to be very similar to the one obtained when using strategy VSM-PLL. This is only true if the wash-out filter time constant ($T_{WD}$) is high enough. \color{black}

\subsection{Strategy 4: IP Control}\label{sec:Theo_GFM_VSC_IP_control}
In the IP control scheme (Fig.~\ref{fig:VSM_configurations}-(d)), when a fault occurs, the swing equation of the \ac{GFVSC} (in the Laplace domain) is given by:
\begin{eqnarray}\label{eq:IP_swing_eq}
    \frac{1}{2sH_{GFM}}(p_{g}^{0}-\cancelto{0}{p_{g}}) - K_{P}\cancelto{0}{p_{g}} = \omega	
\end{eqnarray}
Hence,
\begin{equation}\label{eq:IP_swing_eq_Pg_zero}
\frac{d \omega}{dt} = \frac{1}{2 H_{GFM}}p_{g}^{0}
\end{equation}
This means that the frequency increase during the fault when using IP control is higher than the one obtained in VSM with PLL. This is not trivial, since the behaviour of strategies VSM-PLL and IP control are equivalent under small disturbances, if they are properly tuned (see Section~\ref{sec:sscs_equivalences}). Nevertheless, transient-stability behaviour is different when using IP control and the reason is that during the fault the active-power injection of the ($p_g$) goes to zero and this produces a higher acceleration of the frequency, according to the block diagram of this control strategy (Fig.~\ref{fig:VSM_configurations}-(d)). Therefore, the CCT obtained with IP control is expected to the lower than the one obtained with strategy VSM with PLL.

\subsubsection{Conclusion}\label{sec:Theo_GFM_VSC_conclusion}
According to the theoretical analysis, the highest CCTs are achieved with strategies VSM-PLL and VSM-Washout. The CCT obtained with the strategy of VSM without PLL is expected to be lower, due to the fact that it has a lower value of the damping coefficient ($D_{GFM}=20-25$~pu), because the damping term also plays the role of primary frequency controller. The CCT with IP control is expected to the lower than the one obtained with strategies VSM-PLL and VSM-Washout, because during the fault the active-power injection of the ($p_g$) goes to zero and this produces a higher acceleration of the frequency in the former, due to its block diagram. Qualitative analysis cannot provide conclusions about which CCT is higher: with VSM-noPLL or with IP control.  If the the IP control is tuned to have a high damping ratio, it is expected to have a better performance than when using strategy VSM without PLL, due to the behaviour of the \ac{GFVSC} after the fault clearing. However, which of these two strategies will have higher CCTs would depend on the control tuning and scenarios considered. \color{black}

\subsection{Virtual active-power control (VAPC)}\label{sec:Theo_VAPC}
When using VAPC (Fig.~\ref{fig:self-synchronisation_control_Pvirt_sscs} of Section~\ref{sec:sscs_VAPC_Controller}), the unsaturated virtual active power, $p_{g}^{virt}$, is used as feedback signal in the \ac{GFVSC} self-synchronisation strategy. This means that transient stability behaviour of the \ac{GFVSC} is driven by $P-\delta$ curves using $p_{g}^{virt}$, independently that that current limit is reached~\cite{VattaKkuniK2024}, which is precisely the advantage of VAPC. \color{black} The unsaturated virtual active-power injection of the \ac{GFVSC}, as a function of the angle of its modulated voltage is given by~\cite{VattaKkuniK2024,LabaY2023}: 
\begin{equation}\label{eq:GFM_pg_VAPC}
\begin{split}
    p_{g}^{virt} = \left\lbrace\begin{array}{ccc} 
        \frac{e_{m}^{0} v_{e}}{x_{c}+x_{g}} \sin \delta, &\text{if}& \delta < \delta_a \text{ } (i_g < i_g^{\max}) \\\\
        \frac{e_{m}^{0} v_{e}}{x_{c}+x_{g}/K_{CL}} \sin \delta,, &\text{if}& \delta \geq \delta_a \text{ } (i_g = i_g^{\max})
    \end{array}\right.
\end{split}
\end{equation}
where $\delta_a$ is the angle when the current injection of the \ac{GFVSC} reaches its current limit ($i_g = i_g^{\max}$), and the term $K_{CL}$ is defined as:
\begin{eqnarray}\label{eq:EAC_VAPC_KCL}
    K_{CL}=K_{CL}(\delta)=\frac{\bar{i}_{g}^{ref'}}{\bar{i}_{g}^{ref}}
\end{eqnarray}
where $\bar{i}_{g}^{ref'}=i_{g,d}^{ref'} + ji_{g,q}^{ref'}$ is the unsaturated current reference (before current limitation) and $\bar{i}_{g}^{ref}=i_{g,d}^{ref} + ji_{g,q}^{ref}$ is the saturated current reference (after current limitation) (see Fig.~\ref{fig:VSC_L_control_loops_general} of Section~\ref{sec:VSC_V}).
Since the \ac{GFVSC} has a CL-CSA current limiter with equal priority to $d$- and $q$-axis current injections, $\bar{i}_{g}^{ref'}$ and $\bar{i}_{g}^{ref}$ have the same phase; therefore, $K_{CL}$ is a real number. Notice also that $K_{CL} \ge 1$ always.

\FloatBarrier
\begin{figure}[!htbp]
	\begin{center}
		\centering
		\includegraphics[width=0.8\columnwidth]{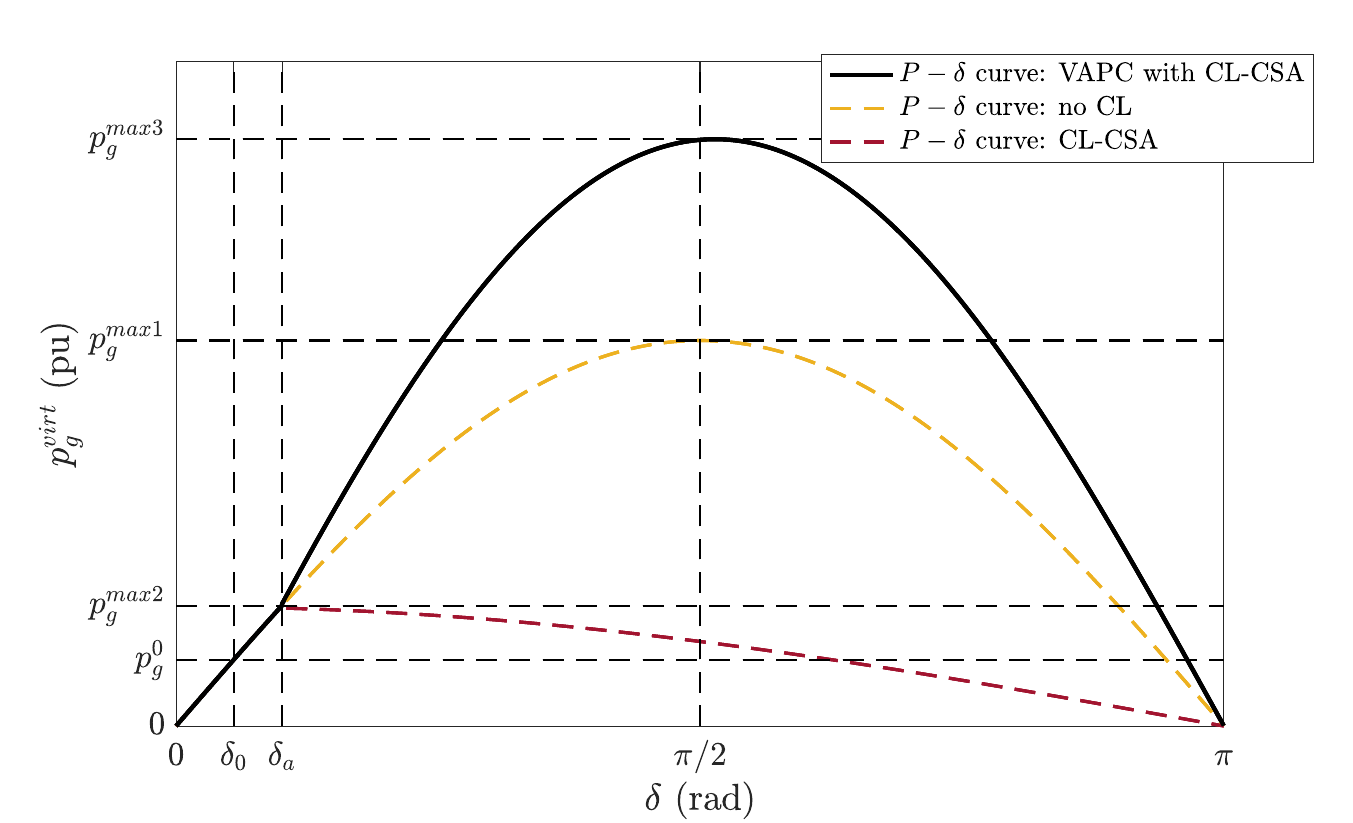}
		\caption{\ac{GFVSC} with Virtual active power control (VAPC) and CL-CSA: $P-\delta$ curve.}
		\label{fig:EAC_case3}
	\end{center}
\end{figure}

\FloatBarrier
Fig.~\ref{fig:EAC_case3} shows the the $P-\delta$ curve of the unsaturated virtual active-power injection of the \ac{GFVSC} ($p_{g}^{virt}$), when using VAPC and current limiter CL-CSA~\cite{VattaKkuniK2024,LabaY2023}. It is clear from Fig.~\ref{fig:EAC_case3} that with VAPC the $P-\delta$ curve is larger \color{black} than the ones obtained with \ac{GFVSC} without CL and with CL-CSA. Therefore, the CCT \color{black} obtained with VAPC will be \color{black} much higher that the ones obtained without VAPC (without CL and with CL-CSA). This is because when the current injection of the \ac{GFVSC} reaches its current limit, the denominator of the sinusoidal expression of $p_{g}^{virt}$~(\ref{eq:GFM_pg_VAPC}) has a lower reactance: $x_{c}+x_{g}/K_{CL}$ with $K_{CL}\ge 1$, as explained in~\cite{VattaKkuniK2024,LabaY2023}. 

As a conclusion, with VAPC the CCT of the fault can be increased significantly and VAPC can be applied to all \ac{GFVSC} self-synchronisation strategies of Section~\ref{sec:sscs_VSM}.

\subsection{Proposed frequency limiter}\label{sec:Theo_frequency_limiter}
The frequency limiter for \acp{GFVSC} proposed in Section~\ref{sec:sscs_freq_lim} is analysed now, using the $P-\delta$ curves of Fig.~\ref{fig:EAC_case2}. The critical clearing angle ($\delta_{cl}^{crit}$) is independent of the \ac{GFVSC} self-synchronisation strategy used and it depends only on the $P-\delta$ curve. However, the proposed frequency limiter will have an impact on the behaviour of the \ac{GFVSC} self-synchronisation strategy and, therefore, on the CCT of the fault. With then proposed frequency limiter, the derivative of the frequency imposed by the \ac{GFVSC} will be given by:
\begin{equation}\label{eq:GFM_VSC_freq_lim_dw_dt}
\begin{split}
    \frac{d \omega}{d t}  \left\lbrace\begin{array}{ccc} 
        = \text{GFM-VSC self-sync. strategy}, &\text{if}& \omega^{min}<\omega < \omega^{max}. \\
        \ge 0 , &\text{if}& \omega = \omega^{min}. \\
        \le 0 , &\text{if}& \omega = \omega^{max}.
    \end{array}\right.
\end{split}
\end{equation}
It is clear that if the frequency is limited, the critical clearing angle will be reached later ($\delta_{cl}^{crit}$) and, therefore, the CCT of the fault will increase significantly. Furthermore, the proposed frequency limiter can be applied to all \ac{GFVSC} self-synchronisation strategies of Section~\ref{sec:sscs_VSM}.

\FloatBarrier
\section{Case study and results}\label{sec:sscs_Results_VSM_IP}
\noindent A 100 MVA \ac{GFVSC}-based generator connected to an infinite grid is considered, as shown in Fig.~\ref{fig:study_sys}. Data of the system can be found in the Appendix. Simulations were carried out with VSC\_Lib tool, an open-source tool based on Matlab + Simulink + SimPowerSystems developed by L2EP-LILLE in Migrate project~\cite{L2EP_VSC_GF_2020,MIGRATE_WP3_2018,Qoria2019a}.
\begin{figure}[!htbp]
\begin{center}
\includegraphics[width=0.8\columnwidth]{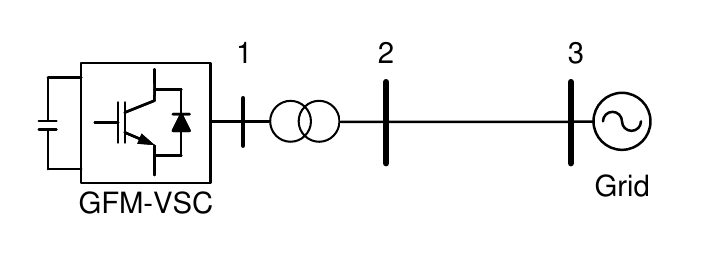}
\caption{\ac{GFVSC} connected to an infinite bus.}
\label{fig:study_sys}
\end{center}
\end{figure}

Transient-stability performance of the following \ac{GFVSC} self-synchronisation strategies will be analysed and compared:
\begin{itemize}
    \item Strategy 1: VSM-noPLL ($D_{GFM}=20$~pu).
    \item Strategy 1: VSM-noPLL ($D_{GFM}=203$~pu) (for illustrative purposes). \color{black}
    \item Strategy 2: VSM-PLL ($D_{GFM}=203$~pu).
    \item Strategy 3: VSM-Washout ($D_{GFM}=203$~pu, $T_{WD}=2$~s).
    \item Strategy 4: IP control ($K_p=0.0096$~pu).
\end{itemize}

The design considerations and values of the parameters are depicted in Table \ref{tab:GFM_VSC_self_sync_design_results}. Notice that in the case of VSM without PLL, the damping coefficient plays also the role of frequency droop and a suitable proportional gain for this purpose must be used: $D_{GFM}=20$~pu, which results in a low damping ratio. An additional case of VSM without PLL is considered using the same design criteria than the other two VSM-type strategies, but it is just for comparison purposes (VSM-noPLL with $D_{GFM}=203$~pu).

\begin{table}[htbp]
    \caption{Design and parameters of GFM-VSC self-synchronisation strategies.} 
    \begin{center}
    \scalebox{0.85}{
    \begin{tabular}{|l|c|c|}
    \hline
    \multirow{2}{*}{\textbf{Strategy}} &  Specified & Calculated \\ 
    & values & value  \\ 
    \hline
     \multirow{4}{*}{Strategy 1: VSM-noPLL} & $H_{GFM}=5$ s, $D_{GFM}=20$ pu&    \\
     & PFR (implicit):  &  $\zeta=0.0691$ \\
     & $K_{PFR}=D_{GFM}=20$ pu& $\omega_n=14.47$~rad/s  \\ 
     & $T_{PFR}=0$ s &   \\  \hline
     \multirow{2}{*}{Strategy 2: VSM-PLL} & $H_{GFM}=5$ s, $\zeta=0.7$ & $D_{GFM}=203$ pu \\
     & PFR: $K_{PFR}=20$ pu, $T_{PFR}=1$ s  & $\omega_n=14.47$~rad/s  \\ \hline
     \multirow{3}{*}{Strategy 3: VSM-Washout}& $H_{GFM}=5$ s, $\zeta=0.7$ & $D_{GFM}=203$ pu  \\
     & PFR: $K_{PFR}=20$ pu, $T_{PFR}=1$ s & $\omega_n=14.47$~rad/s  \\ 
     & Wash-out: $T_{WD}=2$ s &  \\
    \hline
    \multirow{2}{*}{Strategy 4: IP control} & $H_{GFM}=5$ s, $\zeta=0.7$ & $K_P=0.0096$ pu  \\
     & PFR: $K_{PFR}=20$ pu, $T_{PFR}=1$ s & $\omega_n=14.47$~rad/s  \\ \hline
    \end{tabular}
    }
    \label{tab:GFM_VSC_self_sync_design_results}
    \end{center}
    \end{table}

In Subsection~\ref{sec:sscs_Results_fault}, non-linear time-domain simulations are carried out to analyse the performance of the \ac{GFVSC} self-synchronisation strategies when a fault occurs. Subsection~\ref{sec:sscs_Results_CCTs} compares the critical clearing times obtained with each \ac{GFVSC} self-synchronisation strategy. Subsection~\ref{sec:sscs_VAPC_FLC_CCTs} analyses the performance of VAPC and the proposed FLC, when they are applied to the different \ac{GFVSC} self-synchronisation strategies. Finally, Subsection~\ref{sec:sscs_GFM_VSC_VSM_Washout_CCTs} analyses the impact on transient stability of the wash-out-filter time constant in Strategy 3 (VSM-Washout).
\color{black}

\subsection{Short-circuit simulation}\label{sec:sscs_Results_fault}
\begin{figure}[!htbp]
\begin{center}
\includegraphics[width=0.7\columnwidth]{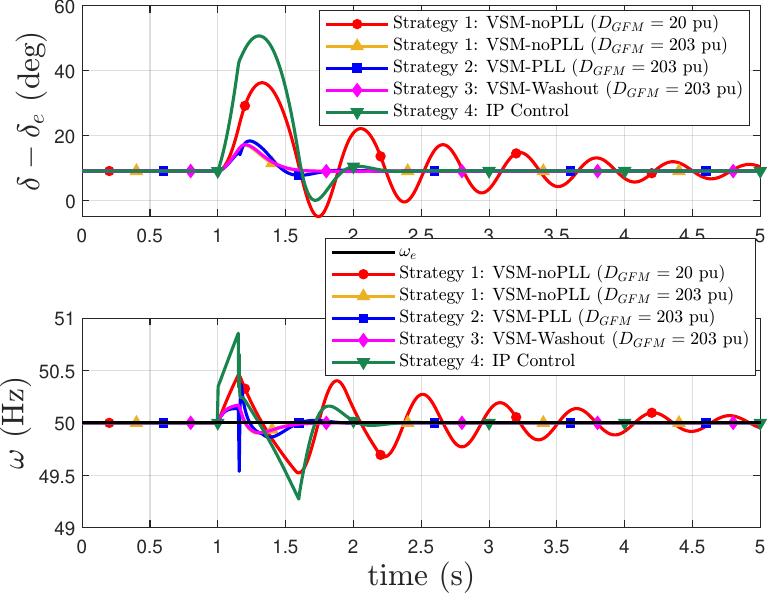}
\caption{Fault 1 cleared after 150 ms. (top) Angle difference and (bottom) Frequency of the \ac{GFVSC}.}
\label{fig:AnglFrqc_PFR_Fault}
\end{center}
\end{figure}

\noindent A three-phase-to-ground short circuit is applied to the line 2-3 of the test system of Fig.~\ref{fig:study_sys} \color{black} at $t=1$ s, close to bus 2, and it is cleared 150 ms later (Fault 1, for short). \color{black} Fig.~\ref{fig:AnglFrqc_PFR_Fault} shows the angle difference between the \ac{GFVSC} and the grid ($\delta-\delta_e$) as well as the frequency deviation of the VSC with respect to the grid frequency. Synchronism is maintained in all cases, although the IP controller showed the largest angle and frequency increase during the transient. After the fault clearing, the VSM-noPLL using $D_{GFM}=20$~pu exhibits oscillations before reaching the steady state due to its low damping factor value. However, VSM-PLL and VSM-Washout (and VSM-noPLL, but with $D_{GFM}=203$~pu) present the smallest variation of the angle difference and the frequency of the \ac{GFVSC} during the transient, producing the best results in terms of transient stability. 

\begin{figure}[!htbp]
\begin{center}
\includegraphics[width=0.7\columnwidth]{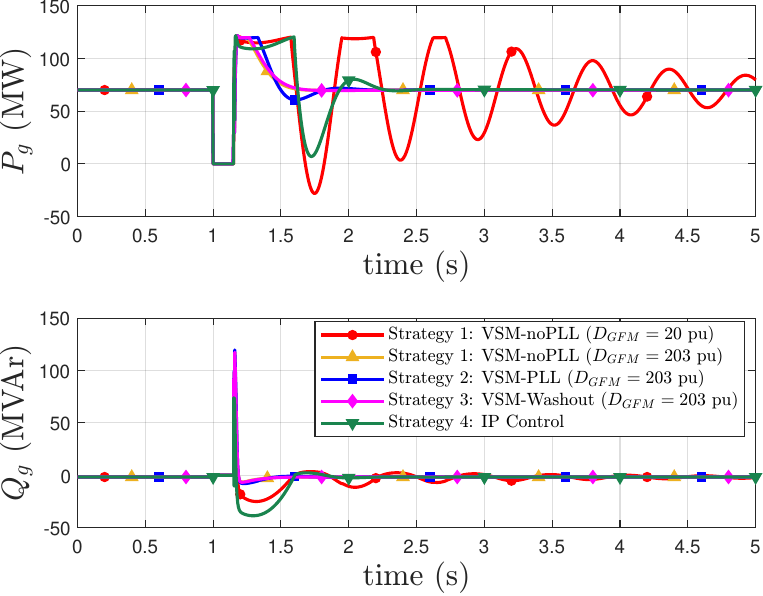}
\caption{Fault 1 cleared after 150 ms. (top) Active and (bottom) Reactive of the \ac{GFVSC}.}
\label{fig:AR_Powers_PFR_Fault}
\end{center}
\end{figure}

\begin{figure}[!htbp]
\begin{center}
\includegraphics[width=0.7\columnwidth]{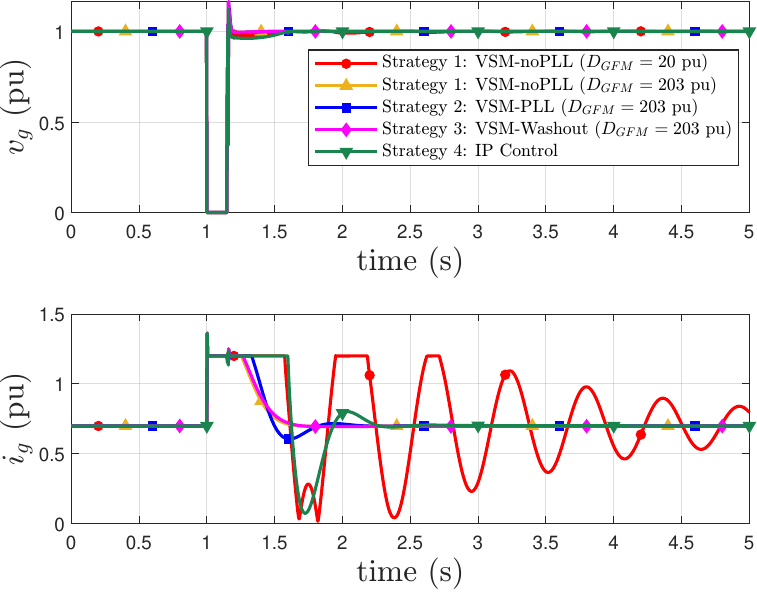}
\caption{Fault 1 cleared after 150 ms. (top) voltage at the terminal and (bottom) Current injection of the \ac{GFVSC}.}
\label{fig:Current_PFR_Fault}
\end{center}
\end{figure}

Fig.~\ref{fig:AR_Powers_PFR_Fault} displays the GFM-VSC's active and reactive power injections, and Fig.~\ref{fig:Current_PFR_Fault} shows the voltages and the current injections of the VSC. During the fault, in all self-synchronisation strategies the active power injection experiences a sharp drop, triggering the converter current limiter (see Fig.~\ref{fig:Current_PFR_Fault}) and accelerating the converter. Once the fault is cleared, the active power injection sharply rises to the limit set by the current limiter ($i_{s}^{max} = 1.25$~pu). Finally, the \ac{GFVSC} controls its modulated voltage according to Fig.~\ref{fig:VSC_L_control_loops_general}, and its reactive-power injection changes accordingly.   \color{black}

A question that arises is, if the behaviour of VSM with PLL and the IP controller is theoretically equivalent (under small disturbances), why do they exhibit so different behaviour under large disturbances, such as short circuits. In all strategies, during the fault, the converter output power goes to zero ($p_{g}=0$). In the IP controller, this leads to the acceleration of the GFM-VSC as according to equation~(\ref{eq:IP_swing_eq_Pg_zero}) (see Fig.~\ref{fig:VSM_configurations}-(c)). \color{black} Hence, the frequency change of the GFM-VSC during the fault is higher, being closer to loss of synchronism.

\begin{figure}[!htbp]
\begin{center}
\includegraphics[width=0.7\columnwidth]{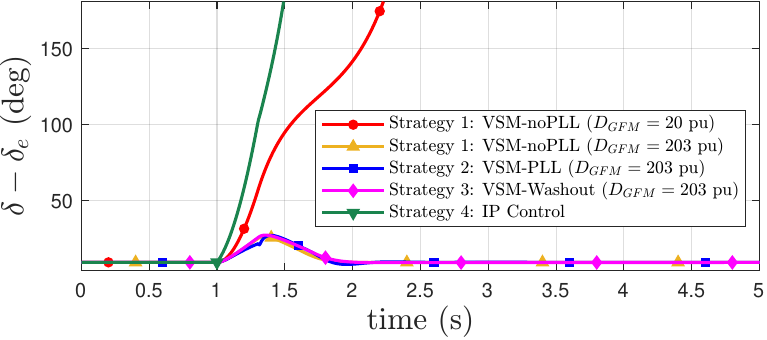}
\caption{Fault 1 cleared after 300 ms: Angle difference of the \ac{GFVSC}.}
\label{fig:Angle_PFR_Fault300}
\end{center}
\end{figure}

Finally, Fig.~\ref{fig:Angle_PFR_Fault300} shows the angle difference between the \ac{GFVSC} and the grid when the clearing time of Fault 1 is 300~ms. Synchronism is lost when the \ac{GFVSC} uses VSM-noPLL (with $D_{GFM}=20$ pu) and IP control, while the system is stable when the \ac{GFVSC} uses  VSM-noPLL (with $D_{GFM}=203$ pu), with VSM-PLL and with VSM-Washout, confirming the conclusions drawn previously.    \color{black}

\FloatBarrier
\subsection{Critical clearing times}\label{sec:sscs_Results_CCTs}
\noindent The Critical Clearing Time (CCT) of a fault is defined as the maximum duration that it can have without provoking loss of synchronism, and it is normally used as an indicator for transient-stability margin. 
Table~\ref{tab:GFM_VSC_CCTs} presents the critical clearing times (CCTs), in milliseconds (ms)\color{black}, of the fault simulated in previous subsection (Fault 1, \color{black} applied to line 2-3, close to bus 2, of the system of Fig.~\ref{fig:study_sys}) obtained for the different \ac{GFVSC} self-synchronisation strategies. The CCT evaluation takes into account the presence or absence of two additional mechanisms: the proposed frequency limiter (FLC) and the virtual active power control (VAPC). 

Strategy 1 (VSM-noPLL) was evaluated with two different values of the damping parameter ($D_{GFM}=20$ and $D_{GFM}~=~203$~pu). 

Strategies \color{black}VSM-noPLL with $D_{GFM}=203$~pu, and VSM-PLL using the same damping coefficient, exhibit the largest CCTs (1600~ms and 1700~ms, respectively), while VSM-noPLL but with $D_{GFM}=20$~pu has a smaller CCT (280 ms), indicating a poorer transient-stability performance. Furthermore, the IP-control strategy has the lowest CCT (190 ms), meaning the worst transient-stability margin. Although the IP controller is similar to the VSM-PLL strategy in small signal analysis, it differs under large disturbances, due to the fact that the active-power injection goes close to zero during the fault. This occurs \color{black} at the frequency change of the GFM-VSC is higher than the one obtained with other self-synchronisation strategies, as discussed in previous subsection.

\begin{table}[!htbp]
    \centering
    \caption{Critical clearing times (CCTs). }
    \scalebox{0.85}{
    \begin{tabular}{|l|c|}
        \hline
        \textbf{\ac{GFVSC} control strategy}     & \textbf{CCT (ms)}   \\ \hline
        Strategy 1: VSM-noPLL ($D_{GFM}=20$~pu)     & 280   \\
        Strategy 1: VSM-noPLL ($D_{GFM}=203$~pu)    & 1600  \\
        Strategy 2: VSM-PLL ($D_{GFM}=203$~pu)      & 1700  \\
        Strategy 3: VSM-Washout ($D_{GFM}=203$~pu)  & 1020  \\
        Strategy 4: IP control                      & 190   \\ \hline
    \end{tabular}
    }
    \label{tab:GFM_VSC_CCTs}
\end{table}

\FloatBarrier
\subsection{Performance of virtual active power control (VAPC) and proposed frequency (FLC) limiter}\label{sec:sscs_VAPC_FLC_CCTs}

\begin{figure}[!htbp]
\begin{center}
\includegraphics[width=0.7\columnwidth]{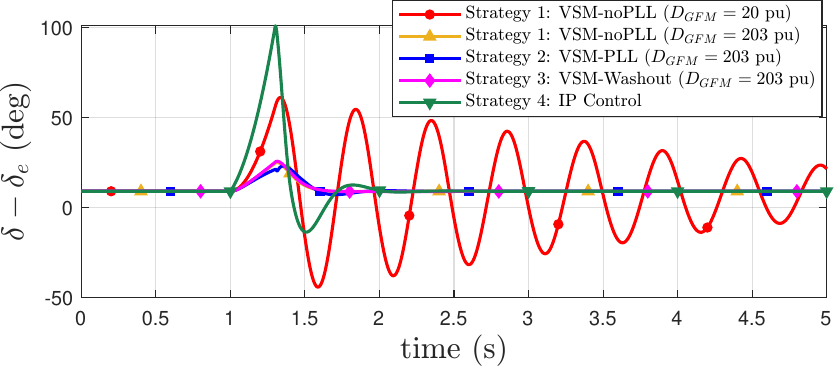}
\caption{GFM-VSC with VAPC. Fault 1 cleared after 300 ms: Angle difference of the \ac{GFVSC}.}
\label{fig:Angle_Pv_PFR_Fault300}
\end{center}
\end{figure}

Fault 1 with a clearing time of 300 ms is simulated now. Fig.~\ref{fig:Angle_Pv_PFR_Fault300} shows the angle difference of the \ac{GFVSC} when VAPC is implemented (Section \ref{sec:sscs_VAPC_Controller}). VAPC concept improves transient stability when it is applied to all \ac{GFVSC} self-synchronisation strategies, since synchronism is maintained in all cases (see Fig. \ref{fig:Angle_PFR_Fault300} for comparison).  \color{black}

Fault 1 with a clearing time of 300 ms is simulated now. Fig.~\ref{fig:Angle_Pv_PFR_Fault300} shows the angle difference of the \ac{GFVSC} when the proposed frequency limitation algorithm (FLC) is implemented  (Section \ref{sec:sscs_freq_lim}), also for Fault 1 with a clearing time of 300 ms.  The frequency of the \ac{GFVSC} is also showed, to understand the frequency limitation process produced by FLC. Parameters of Fig. \ref{fig:frequencylimiter_sscs} and Eq.\color{black}~(\ref{eq_omega_max_min}) are set to $\Delta \omega_{i}^{max}=0.005$ pu, $v_{A,i}=0.5$ pu and $v_{B,i}=0.9$. The proposed frequency limiter (FCL) also improves transient stability when it is applied to all \ac{GFVSC} self-synchronisation strategies, because synchronism is maintained in all cases. Notice that the frequency of the GFM-VSC is limited during the fault (Fig.~\ref{fig:Angle_Pv_PFR_Fault300}), preventing loss of synchronism.  \color{black}

\begin{figure}[!htbp]
\begin{center}
\includegraphics[width=0.7\columnwidth]{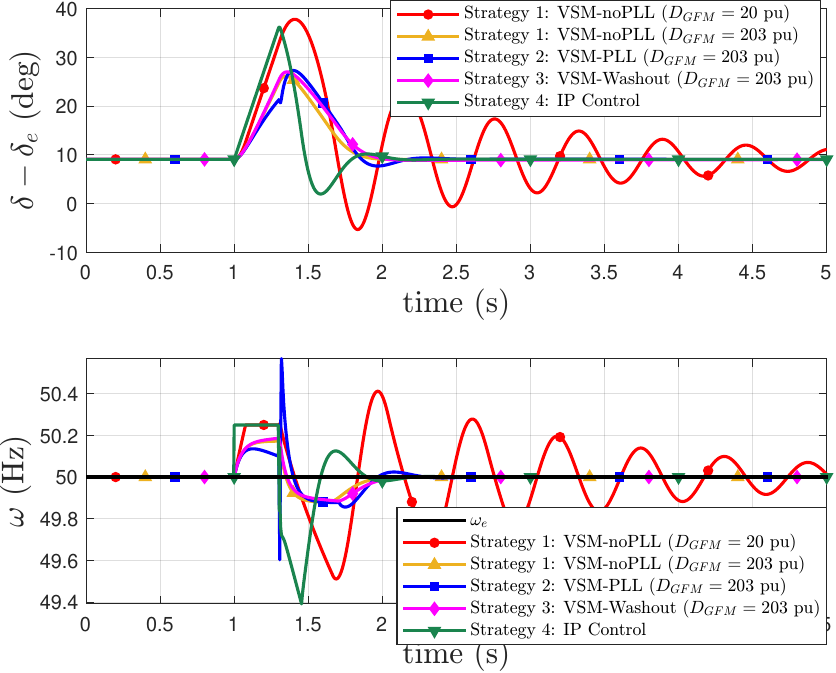}
\caption{GFM-VSC with proposed FLC. Fault 1 cleared after 300 ms. (top) Angle difference and (bottom) Frequency of the \ac{GFVSC}.}
\label{fig:AnglFrqc_FLC_PFR_Fault300}
\end{center}
\end{figure}

\FloatBarrier

Finally, Table~\ref{tab:CCTs_vapc_flc} shows the CCTs of Fault 1 under different combinations of FLC and VAPC activation or deactivation:
\begin{itemize}
    \item \textbf{No-FLC/No-VAPC} (Base case, second column in Table~\ref{tab:CCTs_vapc_flc}): Neither the FLC nor VAPC is used.
    \item \textbf{No-FLC/VAPC} (third column in Table~\ref{tab:CCTs_vapc_flc}): VAPC is used only.
    \item \textbf{FLC/No-VAPC} (forth column in Table~\ref{tab:CCTs_vapc_flc}): With FLC only.
    \item \textbf{FLC/VAPC} (fifth column in Table~\ref{tab:CCTs_vapc_flc}): Both mechanisms are activated. 
\end{itemize}

\begin{table}[!htbp]
    \centering
    \caption{Critical clearing times (CCTs) with and without the proposed frequency limiter (FLC) and virtual active power control (VAPC).}
    \scalebox{0.85}{
    \begin{tabular}{|l|cc|cc|}
        \cline{2-5}
        \multicolumn{1}{c}{} & \multicolumn{4}{|c|}{\textbf{CCT} (ms)}\\
        \hline
        \multirow{2}{*}{\textbf{\ac{GFVSC} control strategy}} & \multicolumn{2}{c|}{\textbf{No-FLC}} & \multicolumn{2}{c|}{\textbf{FLC}} \\
        ~ & \textbf{No-VAPC} & \textbf{VAPC} & \textbf{No-VAPC} & \textbf{VAPC} \\ \hline
        Strategy 1: VSM-noPLL ($D_{GFM}=20$~pu)     & 280  & 540   & 940  & 1790 \\
        Strategy 1: VSM-noPLL ($D_{GFM}=203$~pu)    & 1600 & 2640  & 1600 & 2640 \\
        Strategy 2: VSM-PLL ($D_{GFM}=203$~pu)      & 1700 & 3440  & 1720 & 3830 \\
        Strategy 3: VSM-Washout ($D_{GFM}=203$~pu)  & 1020 & 1900  & 1020 & 1970 \\
        Strategy 4: IP control                      & 190  & 390   & 1160 & 1830 \\ \hline
    \end{tabular}
    }
    \label{tab:CCTs_vapc_flc}
\end{table}

The third column in Table~\ref{tab:CCTs_vapc_flc} (No-FLC/VAPC) demonstrates significant improvements in the CCTs when using VAPC. Meanwhile columns 4 and 5 (FLC/No-VAPC and FLC/VAPC, respectively) show the results by applying the proposed frequency limiter for GFM-VSCs\color{black}. The proposed frequency limiter also results in significant improvements in CCTs in all the GFM-VSC self-synchronisation strategies. 

VSM-noPLL strategy with the lowest damping factor ($D_{GFM}=20$~pu) shows a significant increase in CCT when FLC and/or VAPC are applied. VSM-noPLL experiences a notable increase in CCT from 280 ms (No-FLC/No-VAPC) to 940 ms when the frequency limiter is enabled (FLC/No-VAPC), showcasing its effectiveness in improving stability. CCTs reach 1790 ms when both FLC and VAPC are enabled (FLC/VAPC), compared to the base case. 

Significant improvements have also been made to the IP-control strategy, which benefits considerably from the use of FLC and VAPC. CCT increases from 190 ms without the mechanisms to 1830 ms with both enabled.

As a conclusion, the concept of VAPC and the proposed FLC can be applied to all GFM-VSC self-synchronisation strategies analysed and they are effective to improve transient stability in all cases. Furthermore, VAPC and FLC can be used together, obtaining further improvements.

\FloatBarrier
\subsection{Strategy 3: VSM-Washout - Effect of the wash-out filter time constant}\label{sec:sscs_GFM_VSC_VSM_Washout_CCTs}

In Strategy 3 (VSM-Washout), the time constant of the wash-out filter ($T_{WD}$) has an impact of the performance of Strategy 3 (VSM-Washout) and it is analysed now. A summary of the guidelines for the design of the wash-out filter time constant is as follows (Subsection~\ref{sec:sscs_equivalences}):
\begin{itemize}
    \item Parameter $T_{WD}$ should be relatively high, in order to avoid interactions and achieve a good performance in terms of small-signal stability and transient stability. Reference~\cite{Xiong2021} proposes a curve for feasible regions of parameters $D_{GFM}$ and $T_{WD}$ (see Fig.~\ref{fig:VSM_configurations}-(c)), in terms of transient stability, concluding that concluding that $T_{WD}\ge 1$~s could be a reasonable value for typical parameters. 
    \item Parameter $T_{WD}$ should not be excessively high, to avoid active-power changes during very-slow fluctuations.
    \item Then, the guidelines for the design of the GFM-VSC self-synchronisation strategies can be used (Table~\ref{tab:GFM_VSC_self_sync_design} of Subsection~\ref{sec:sscs_equivalences})), assuming a fixed value of $T_{WD}$. Finally, the behaviour should be checked also by time-domain simulation.
\end{itemize}

Taking the aspects above into account, the choice of the wash-out filter time constant for Strategy 3 used in this work was $T_{WD}=2$~s.

\begin{figure}[!htbp]
\begin{center}
\includegraphics[width=0.7\columnwidth]{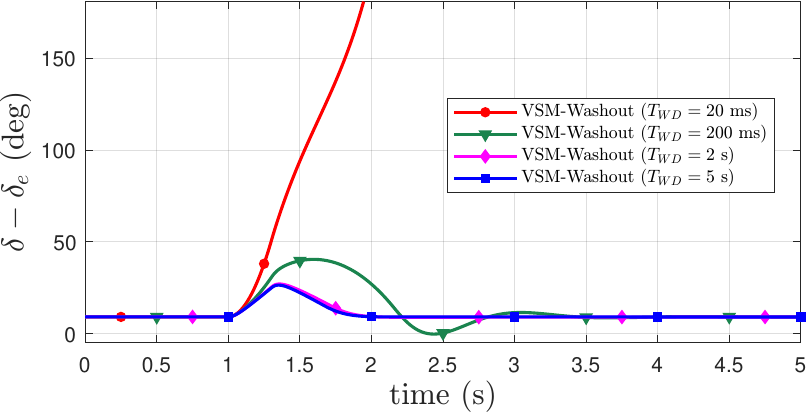}
\caption{Fault 1 cleared after 300 ms: Angle difference of the \ac{GFVSC}.}
\label{fig:VSM_TWD_Angle_WO_TWD_PFR_Fault300}
\end{center}
\end{figure}

The impact of the wash-out filter in Strategy 3 is analysed now for different values of $T_{WD}$: $T_{WD}=20$~ms, $T_{WD}=20$~ms, $T_{WD}=2$~s and $T_{WD}=5$~s. Fig.~\ref{fig:VSM_TWD_Angle_WO_TWD_PFR_Fault300} shows the angle difference of the GFM-VSC and the infinite grid, when Fault 1 with a clearing time of 300~ms is simulated. When $T_{WD}=20$~ms, synchronism is lost. When $T_{WD}=200$~ms, synchronism is maintained, but the angle excursions are high. For $T_{WD}=2$~s and $T_{WD}=5$~s, synchronism is maintained and angle excursions are much lower than in the other cases. 

In strategy 3 (VSM-Washout), transient-stability behaviour should be also taken into account. In previous work, the design of the wash-out filter time constant in strategy VSM-Washout has not been addressed explicitly. However, reference~\cite{Xiong2021} has analysed the impact of the wash-out filter in the so-called transient damping method (TDM), and \color{black} conclusions can be applied directly to the VSM-Washout strategy. The work in~\cite{Xiong2021} proposes a curve for feasible regions of parameters $D_{GFM}$ and $T_{WD}$, in terms of transient stability. Mainly,in order to use high values of $D_{GFM}$, the wash-out filter time constant should be high enough, concluding that $T_{WD}\ge 1$ s could be a reasonable value, for the test system analysed. In addition, in Strategy 3 the wash-out filter add a state variable and the closed-loop system dynamics is more complex and it is not a second-order system any more. However, this aspect can be addressed by using a high enough value of $T_{WD}$ and it is recommended to check the results of the design by time-domain simulation. Finally, notice that \color{black} the wash-out filter time constant $T_{WD}$ should not be excessively high, because it would not be able to avoid active-power changes under frequency offsets, which is its role. \color{black}

Finally, \ref{tab:GFM_VSC_VSM_Washout_CCTs} depicts the CCTs of fault 1 when Strategy 3 is used, for the different values of the washout filter time constant. The highest CCTs are obtained with $T_{WD}=2$ s and $T_{WD}=5$ s, proving clearly that the best results are obtained with these values. When $T_{WD}=0.20$~ms, the system is stable and it is unstable even for small perturbations, that is why the CCT is 0~ms. This proves that too low values of $T_{WD}$ will produce poor results.

\color{black}

\begin{table}[!htbp]
    \centering
    \caption{CCTs of VSM-Washout strategy with different time constants.}
    \scalebox{0.85}{
    \begin{tabular}{|l|c|}
        \hline
        \textbf{\ac{GFVSC} control strategy}     & \textbf{CCT (ms)}   \\ \hline
        Strategy 3: VSM-Washout ($T_{WD}=0.02$~s)  & 0 \\ 
        Strategy 3: VSM-Washout ($T_{WD}=0.2$~s)  & 460  \\
        Strategy 3: VSM-Washout ($T_{WD}=2$~s)  & 1020  \\
        Strategy 3: VSM-Washout ($T_{WD}=5$~s)  & 1290   \\ \hline
    \end{tabular}
    }
    \label{tab:GFM_VSC_VSM_Washout_CCTs}
\end{table}

\color{black}
\FloatBarrier
\section{Conclusions}\label{sec:sscs_conclusion}
\noindent This paper analysed and compared transient-stability performance of various self\hyp{}synchronisation strategies for \acp{GFVSC}: VSM-noPLL, VSM-PLL, VSM-Washout and IP control. The paper also analysed different methods to be applied to \ac{GFVSC} self-synchronisation strategies, aimed to improve their transient-stability performance: the concept of virtual unsaturated active-power controller (VAPC), proposed in previous work, and an algorithm for frequency limitation in the \acp{GFVSC} (FLC), which was proposed in this paper.  The conclusions obtained in this paper can be summarised as follows:
\begin{itemize}
    \item The four self-synchronisation mechanisms for \acp{GFVSC} analysed in this work (VSM-noPLL, VSM-PLL, VSM-Washout and IP control) can have a similar behaviour under small disturbances by proper tuning of the parameters. However, they can have different behaviour under large disturbances (transient stability), due to non-linearities.
    \item Strategies VSM-PLL and VSM-Washout produce the best results in terms of transient stability. 
    \item Strategies VSM-noPLL and IP control produce the lowest critical clearing times (CCTs). 
    \item Strategy VSM-noPLL ($D_{GFM}=20$~pu) produces worse results than other strategies in terms on transient stability not due to its control structure, but due to the design of the parameters. Since the damping coefficient used in the self-synchronisation strategy plays also the role of primary frequency controller, it should be designed with typical values for primary-frequency controller proportional gains, which produces a dynamic response of the \ac{GFVSC} with lower damping ratio than with other options, and this jeopardises transient stability. 
    \item The IP control produces worse results than other options in therms of transient stability due to the active-power signal which is subtracted after the inertial-response term. When a fault occurs, the voltage at the PCC becomes close to zero; therefore, the active-power injection becomes close to zero too, producing a high variation of the frequency of the \ac{GFVSC} during the fault, being more vulnerable to loss of synchronism.
    \item The concept of VAPC (proposed in previous work) can be applied to all \ac{GFVSC} self-synchronisation strategies and it improves transient stability in all cases.
    \item The proposed algorithm for frequency limitation (FLC) can be applied to all \ac{GFVSC} self-synchronisation strategies and it improves transient stability in all cases.
    \item Applying Virtual Active Power Control (VAPC) or the proposed Frequency Limitation Control (FLC) together significantly improves transient stability of the four \ac{GFVSC} self-synchronisation strategies analysed, increasing the CCTs. \color{black} 
    \item In VSM-Washout strategy, the wash-out-filter time constant should be high enough to produce good results in terms of transient stability. Reasonable values could be in the range $T_{WD}= 1-10$ s, for the test system considered in this paper. \color{black}
\end{itemize}

\section*{Acknowledgements}
\noindent Work supported by the Spanish Government under a research project ref. PRE2019-088084 and RETOS Project  Ref. RTI2018-098865-B-C31 (MCI/AEI/FEDER, UE); and by Madrid Regional Government under PROMINT-CM Project  Ref. S2018/EMT-4366.

\section*{Contact information of the authors}
\noindent Régulo E. Ávila-Martínez: regulo.avila@iit.comillas.edu, Xavier Guillaud: xavier.guillaud@centralelille.fr, Javier Renedo: javier.renedo@ieee.org, Luis Rouco: luis.rouco@iit.comillas.edu, Aurelio Garcia-Cerrada: aurelio@iit.comillas.edu, Lukas Sigrist: lukas.sigrist@iit.comillas.edu.

\section*{Appendix: data}\label{app:GFM_VSC_data}
\noindent Data of GFM-VSCs are provided in Table~\ref{tab:VSC_parameters_ib_app}. Nominal voltage and frequency of the system are 220~kV and 50~Hz, respectively. Parameters of line 2-3 are $r_g~=~0.01$~pu, $x_g~=~0.1$~pu (base values for pu: 220~kV and 100~MVA). Initial operating point of the GFM-VSC is $P_{g}^0=70$ MW and $v_g^0=1$ pu. Constant modulated voltage set-point term was set to $e_{m}^0=1.0057$ pu.

When using the frequency limiter (FCL) proposed in Section~\ref{sec:sscs_freq_lim}, parameters of Fig. \ref{fig:frequencylimiter_sscs} and (\ref{eq_omega_max_min}) are set to $\Delta \omega_{i}^{max}=0.005$ pu, $v_{A,i}=0.5$ pu and $v_{B,i}=0.9$. \color{black}

\begin{table}[H]
    \caption{Parameters of the GFM-VSCs} 
    \begin{center}
    \begin{tabular}{|l|c|}
    \hline
    \textbf{Parameters} &  \textbf{Values}   \\ 
    VSC's rating are base values for pu& \\ 
    \hline
    Rating VSC, DC voltage, AC voltage & 100 MVA, $ 640$ kV, $400$ kV   \\
    Max. current & 1.20 pu (equal priority for $d-q$ axes)  \\
    Series resistance ($r_{c,i}$) / reactance ($x_{c,i}$) & 0.005 pu / 0.15 pu \\
    (series reactor + transformer) &  \\
    (transformer: 100 MVA 400/320 kV) &  \\
    Current control bandwidth ($\omega_{n,CC}$) &  1500 rad/s \\
    Current control prop. / int. gains ($K_{C,P,i}$ / $K_{C,I,i}$) &  1.0027 pu / 1074.3 pu/s \\
    Quasi-static model, virt. reactance  ($x_v$ / $x_{ev}$) &  0 pu / 0.15 pu \\
    Emulated inertia ($H_{GFM,i}$) & 5  s \\
    Primary freq. controller ($K_{PFR,i}$ / $T_{PFR,i}$)  & 20 pu / 1 s \\
    Primary freq. controller limit ($\Delta p_{g,i}^{max}$)  & 1 pu \\
    PLL control bandwidth ($\omega_{n,PLL}$) & 500 rad/s \\
    PLL control porp. / int. gains ($K_{P,PLL}$/$K_{I,PLL}$) & 3.1831 pu / 795.7747 pu/s \\
    PLL frequency saturation ($\Delta \omega_{PLL}^{max}$) & $\pm$ 0.1 pu \\
    \hline
    \end{tabular}
    \label{tab:VSC_parameters_ib_app}
    \end{center}
    \end{table}





\end{document}